 \documentclass[final,times,twocolumn,authoryear]{elsarticle}

\usepackage{amssymb}

\usepackage{hyperref}
\hypersetup{
    colorlinks=true,
    linkcolor=blue,
    filecolor=blue,      
    urlcolor=blue,
    citecolor=blue
    }

\usepackage{amsmath}
\usepackage{booktabs}
\usepackage{subcaption}
\usepackage{xcolor, soul}

\journal{Computers in Biology and Medicine; DOI: 10.1016/j.compbiomed.2024.109242 }

\RequirePackage{color}\definecolor{RED}{rgb}{1,0,0}\definecolor{BLUE}{rgb}{0,0,1} 
\RequirePackage[stable]{footmisc} 
\DeclareOldFontCommand{\sf}{\normalfont\sffamily}{\mathsf} 
\providecommand{\DIFaddbegin}{} 
\providecommand{\DIFaddend}{} 
\providecommand{\DIFdelbegin}{} 
\providecommand{\DIFdelend}{} 
\providecommand{\DIFaddbeginFL}{} 
\providecommand{\DIFaddendFL}{} 
\providecommand{\DIFdelbeginFL}{} 
\providecommand{\DIFdelendFL}{} 
\newcommand{\DIFscaledelfig}{0.5}
\RequirePackage{settobox} 
\RequirePackage{letltxmacro} 
\newsavebox{\DIFdelgraphicsbox} 
\newlength{\DIFdelgraphicswidth} 
\newlength{\DIFdelgraphicsheight} 
\LetLtxMacro{\DIFOincludegraphics}{\includegraphics} 
\newcommand{\DIFaddincludegraphics}[2][]{{\color{blue}\fbox{\DIFOincludegraphics[#1]{#2}}}} 
\newcommand{\DIFdelincludegraphics}[2][]{
\sbox{\DIFdelgraphicsbox}{\DIFOincludegraphics[#1]{#2}}
\settoboxwidth{\DIFdelgraphicswidth}{\DIFdelgraphicsbox} 
\settoboxtotalheight{\DIFdelgraphicsheight}{\DIFdelgraphicsbox} 
\scalebox{\DIFscaledelfig}{
\parbox[b]{\DIFdelgraphicswidth}{\usebox{\DIFdelgraphicsbox}\\[-\baselineskip] \rule{\DIFdelgraphicswidth}{0em}}\llap{\resizebox{\DIFdelgraphicswidth}{\DIFdelgraphicsheight}{
\setlength{\unitlength}{\DIFdelgraphicswidth}
\begin{picture}(1,1)
\thicklines\linethickness{2pt} 
{\color[rgb]{1,0,0}\put(0,0){\framebox(1,1){}}}
{\color[rgb]{1,0,0}\put(0,0){\line( 1,1){1}}}
{\color[rgb]{1,0,0}\put(0,1){\line(1,-1){1}}}
\end{picture}
}\hspace*{3pt}}} 
} 
\LetLtxMacro{\DIFOaddbegin}{\DIFaddbegin} 
\LetLtxMacro{\DIFOaddend}{\DIFaddend} 
\LetLtxMacro{\DIFOdelbegin}{\DIFdelbegin} 
\LetLtxMacro{\DIFOdelend}{\DIFdelend} 
\DeclareRobustCommand{\DIFaddbegin}{\DIFOaddbegin \let\includegraphics\DIFaddincludegraphics} 
\DeclareRobustCommand{\DIFaddend}{\DIFOaddend \let\includegraphics\DIFOincludegraphics} 
\DeclareRobustCommand{\DIFdelbegin}{\DIFOdelbegin \let\includegraphics\DIFdelincludegraphics} 
\DeclareRobustCommand{\DIFdelend}{\DIFOaddend \let\includegraphics\DIFOincludegraphics} 
\LetLtxMacro{\DIFOaddbeginFL}{\DIFaddbeginFL} 
\LetLtxMacro{\DIFOaddendFL}{\DIFaddendFL} 
\LetLtxMacro{\DIFOdelbeginFL}{\DIFdelbeginFL} 
\LetLtxMacro{\DIFOdelendFL}{\DIFdelendFL} 
\DeclareRobustCommand{\DIFaddbeginFL}{\DIFOaddbeginFL \let\includegraphics\DIFaddincludegraphics} 
\DeclareRobustCommand{\DIFaddendFL}{\DIFOaddendFL \let\includegraphics\DIFOincludegraphics} 
\DeclareRobustCommand{\DIFdelbeginFL}{\DIFOdelbeginFL \let\includegraphics\DIFdelincludegraphics} 
\DeclareRobustCommand{\DIFdelendFL}{\DIFOaddendFL \let\includegraphics\DIFOincludegraphics} 
\RequirePackage{listings} 
\RequirePackage{color} 
\lstdefinelanguage{DIFcode}{ 
  moredelim=[il][\color{red}\scriptsize]{\%DIF\ <\ }, 
  moredelim=[il][\color{blue}\sffamily]{\%DIF\ >\ } 
} 
\lstdefinestyle{DIFverbatimstyle}{ 
	language=DIFcode, 
	basicstyle=\ttfamily, 
	columns=fullflexible, 
	keepspaces=true 
} 
\lstnewenvironment{DIFverbatim}{\lstset{style=DIFverbatimstyle}}{} 
\lstnewenvironment{DIFverbatim*}{\lstset{style=DIFverbatimstyle,showspaces=true}}{} 

\begin{document}

\begin{frontmatter}



\title{Less is More:\\ Selective Reduction of CT Data for Self-Supervised Pre-Training of Deep Learning Models with Contrastive Learning Improves Downstream Classification Performance}



\author[1,2]{Daniel Wolf}
\author[1]{Tristan Payer}
\author[2]{Catharina Silvia Lisson}
\author[2]{Christoph Gerhard Lisson}
\author[2]{Meinrad Beer}
\author[2,*]{Michael Götz}
\author[1,*]{Timo Ropinski}

\affiliation[1]{organization={Visual Computing Research Group, Institute of Media Informatics, Ulm University},
            addressline={James-Franck-Ring}, 
            city={Ulm},
            postcode={89081}, 
            country={Germany}}

\affiliation[2]{organization={Experimental Radiology Research Group, Department for Diagnostic and Interventional Radiology, Ulm University Medical Center},
            addressline={Albert Einstein Allee}, 
            city={Ulm},
            postcode={89081}, 
            country={Germany}}

\affiliation[*]{these authors contributed equally to this work}

\begin{abstract}
\textit{Background:} Self-supervised pre-training of deep learning models with contrastive learning is a widely used technique in image analysis. Current findings indicate a strong potential for contrastive pre-training on medical images. However, further research is necessary to incorporate the particular characteristics of these images.

\noindent \textit{Method:}
We hypothesize that the similarity of medical images hinders the success of contrastive learning in the medical imaging domain. To this end, we investigate different strategies based on deep embedding, information theory, and hashing in order to identify and reduce redundancy in medical pre-training datasets. The effect of these different reduction strategies on contrastive learning is evaluated on two pre-training datasets and several downstream classification tasks.

\noindent \textit{Results:}
 In all of our experiments, dataset reduction leads to a considerable performance gain in downstream tasks, e.g., an AUC score improvement from 0.78 to 0.83 for the COVID CT Classification Grand Challenge, 0.97 to 0.98 for the OrganSMNIST Classification Challenge and 0.73 to 0.83 for a brain hemorrhage classification task. Furthermore, pre-training is up to nine times faster due to the dataset reduction. 

\noindent \textit{Conclusions:}
 In conclusion, the proposed approach highlights the importance of dataset quality and provides a transferable approach to improve contrastive pre-training for classification downstream tasks on medical images.

 \noindent \textit{Code:} \url{https://github.com/Wolfda95/Less_is_More}
\end{abstract}




\end{frontmatter}




\section{Introduction}

\begin{figure*}[!ht]
\centering
\includegraphics[width=1\textwidth]{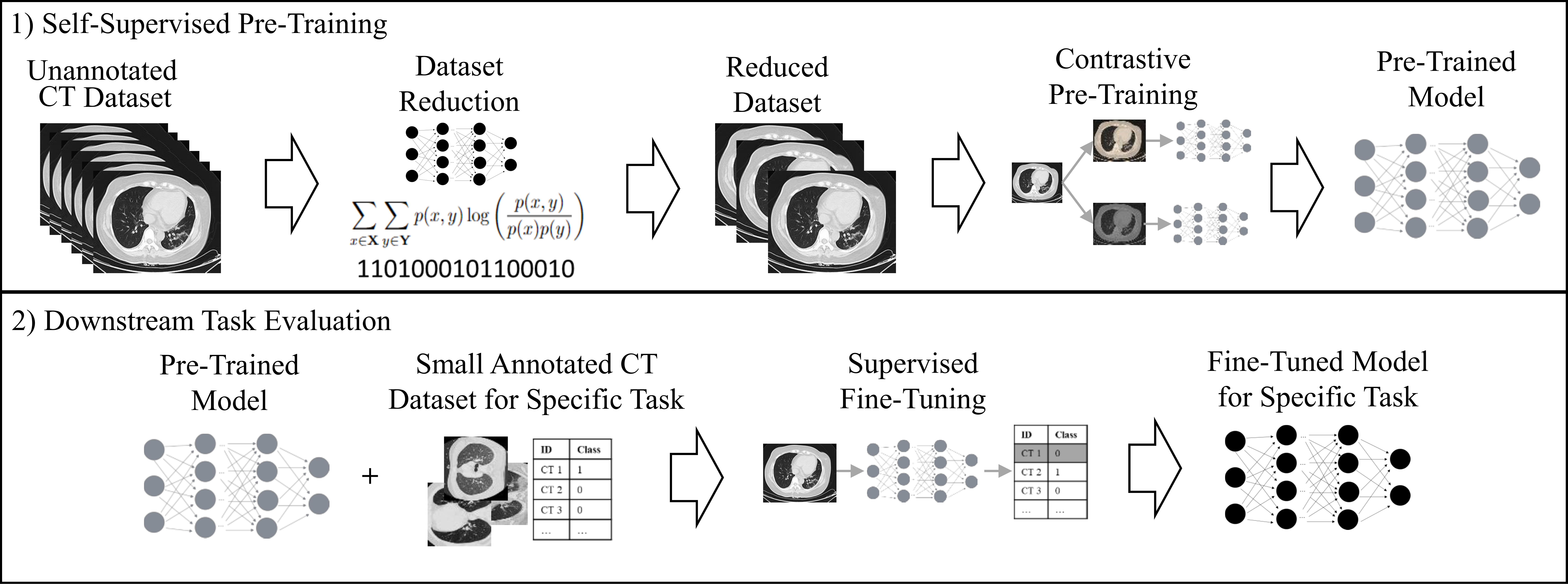}
\caption{This figure gives an overview of our approach to investigate the hypothesis that using all CT slices for contrastive pre-training may lead to performance degradation due to the high similarity of the slices. The first step is to pre-train a deep learning model. Therefore, we start with a dataset of unannotated CT slices, select slices in such a way that we obtain a reduced dataset with increased variation, and pre-train the model with contrastive learning on the reduced dataset. The second step is to evaluate the pre-training on downstream tasks. Therefore, the pre-trained model is fine-tuned with supervised learning on small datasets with annotations for the specific task. Our work compares different strategies to reduce CT image pre-training datasets.}\label{fig:overview}
\end{figure*}

Supervised training of a deep learning model requires large and accurate datasets. Annotations are necessary for all training samples. In the medical imaging domain, annotated datasets for specific tasks are often limited due to factors such as the rarity of diseases, limited access, or the high complexity of annotations~\citep{kiryati2021dataset,maier2018rankings}. To overcome this challenge, deep learning models can be pre-trained on large medical image datasets without annotations, using self-supervised learning techniques~\citep{huang2023self}. These techniques train the models to create meaningful representations from unlabeled datasets, allowing them to learn general high-level features of the images. To fine-tune the models for specific tasks, the so-called ``downstream tasks", small annotated datasets are sufficient after pre-training. Contrastive learning is a state-of-the-art approach for self-supervised pre-training on unannotated images~\citep{balestriero2023cookbook} and according to Huang et al.`s study~\citep{huang2023self}, currently the most popular approach in the medical imaging domain. Several works show remarkable performance gains on medical downstream tasks when the models are pre-trained with contrastive learning on large unannotated medical image datasets compared to training the models from scratch~\citep{ghesu2022contrastive,tang2022self,chen2021momentum,azizi2021big,ewen2021targeted,dufumier2021contrastive}. Despite the great potential of contrastive pre-training techniques in the medical domain, the special characteristics of volumetric radiological images, consisting of many consecutive slices, such as CT, MRI or PET, have not been sufficiently exploited. In our work, we evaluate the composition of the pre-training datasets for contrastive learning on CT scans.

When it comes to deep learning on CT scans in general, there are two widely used approaches, both of which show excellent results on clinically relevant imaging tasks. The first approach is to train on the whole CT volumes using a 3D model~\citep{lisson2022deep,andrearczyk2021overview}, and the second approach is to train on individual slices of the volumes using a 2D model~\citep{jiang2022dynamic,xing2022cs,baghdadi2022automated,wang2021deep}. Each approach has its own advantages. While training a 3D model on volumes enables the model to better capture the 3D properties of the images~\citep{avesta2023comparing}, training a 2D model by using each slice of a volume separately can improve performance on small datasets due to the increased sample size~\citep{zettler2021comparison,kern20212d,bhattacharjee2021comparison}, and reduces the computational cost of training and inference as smaller GPUs are sufficient~\citep{zettler2021comparison,kern20212d,yu20202d,nemoto2020efficacy,avesta2023comparing}. 
For both approaches, there are several publications that investigate self-supervised pre-training with contrastive learning, achieving significant performance gains on several CT image downstream tasks. Tang et al.~\citep{tang2022self} and Dufumier et al.~\citep{dufumier2021contrastive} pre-train 3D models on CT volumes, while Wolf et al.~\citep{wolf2023self}, Ghesu et al.~\citep{ghesu2022contrastive} and Chen et al.~\citep{chen2021momentum}  pre-train 2D models on CT slices. In this study, we chose to conduct our experiments with 2D models because we see that it is critical for deep learning to be globally accessible without the need for powerful GPUs, and the advantages of 2D models for sparse data remain significant even when using pre-trained models, as small annotated datasets are a major challenge in the field of medical imaging.

Contrastive learning involves the following steps: A dataset of unlabeled images is used as a starting point. Random augmentations are applied to generate multiple randomly varied samples of each original image. These are fed into a deep learning model to obtain latent-space representations for each sample. The model always compares two representations and is trained to discriminate whether these are derived from the same original image (referred to as positive pairs) or derived from different original images (referred to as negative pairs). Previous works on contrastive pre-training with CT slices have included as many slices as possible, following the traditional approach of maximizing the pre-training dataset~\citep{ghesu2022contrastive,chen2021momentum,wolf2023self}. In this paper, we hypothesize that using all slices of each CT volume in a dataset for contrastive pre-training may lead to performance degradation. We derive our hypothesis from the fact that CT datasets have very low variance compared to natural image datasets due to the high similarity of the CT slices. This may result in the model being unable to discriminate between positive and negative pairs since the similarity between two augmented versions of a slice might be lower than the similarity between two different slices. Our hypothesis is supported by recent work that provides preliminary evidence that this may be a challenge in contrastive learning. Using ImageNet data, Jing et al.~\citep{jing2022understanding} show that a lower variance of the data distribution than the variance caused by the data augmentation of contrastive learning leads to performance degradation in downstream tasks. Conrad and Narayan~\citep{conrad2021cem500k} show on electron microscopy images that low variance in the pre-training dataset affects downstream task results. 

To investigate our hypothesis that using all CT slices for contrastive pre-training may lead to performance degradation, we explore various strategies based on deep embedding, information theory, and hashing to identify and reduce redundancy in pre-training datasets. Figure~\ref{fig:overview} illustrates our general approach. Starting with a dataset of unannotated CT slices, we perform different reduction strategies and pre-train the models with contrastive learning on the reduced datasets. The pre-trainings are evaluated on downstream tasks by fine-tuning the pre-trained models with supervised learning. We choose two pre-training datasets and three downstream classification tasks, the benchmark task for evaluating self-supervised pre-training~\citep{huang2023self}. 
The outcomes support our hypothesis, as the downstream results improve with our dataset reduction strategies. Furthermore, we investigate which reduction strategy is best suited for CT datasets and what is the optimal threshold that represents the best trade-off between high variation but also a sufficiently large number of samples in the pre-training dataset to achieve the best downstream results.  
Finally, our work provides a ready-to-use model for improving self-supervised pre-training on CT datasets for classification downstream tasks. These findings have the potential to improve the handling of small annotated CT datasets while maintaining low computational costs. The pre-trained models, as well as the ready-to-use code, are available on GitHub: \url{https://github.com/Wolfda95/Less_is_More}

\section{Materials and Methods}
In this section, we explain in detail the methods for investigating our hypothesis that using all slices of each CT volume for contrastive pre-training may lead to performance degradation due to the high similarity of the slices.
We first present strategies for selecting slices of CT volumes to obtain a reduced pre-training dataset with increased variation. This is followed by describing the contrastive pre-training methods and datasets. Finally, we introduce the downstream tasks to evaluate the impact of the reduction strategies on contrastive pre-training.

\subsection{Dataset Reduction} 
We investigate our hypothesis by comparing six approaches for slice selection: two baseline approaches and four similarity-based approaches. The similarity-based approaches perform a pairwise comparison between all slices in a volume. A similarity score is calculated for each slice pair. The pairs are sorted from most similar to most dissimilar. Starting from the most similar pair, one slice is removed from the pairs until either all pairs have similarities below a given threshold or until a given number of slices is left. We incorporate commonly used similarity computation methods from different fields, such as information theory, deep embedding, and hashing, without claiming completeness. The methods we selected are well-established for image comparison and computationally fast, which is necessary due to the large number of pairwise comparisons. 

\textbf{ALL}\newcommand{\all}{ALL}: The first baseline approach follows the current state of the art~\citep{chen2021momentum,ghesu2022contrastive}. All slices are included in the training. 

\textbf{EveryN}\newcommand{\everyn}{EveryN}: The second baseline approach is our baseline reduction method. Here, CT datasets are reduced by using every $n$th  slice of a volume. 

\textbf{SSIM}\newcommand{\ssim}{SSIM}: As our first similarity-based approach, we use the Structural Similarity Index (SSIM)~\citep{wang2004image} from information theory, which is a common similarity measure for images~\citep{wang2009mean}.
It compares the luminance, contrast, and structure of two given images ${\bf x}$ and ${\bf y}$ by the equation  
\begin{equation}
{\rm d}({\bf x},{\bf y})={\frac{(2\mu_{\mathrm x}\mu_{y}+(K_{1} L)^{2})(2\sigma_{xy}+(K_{2} L)^{2})}{\left(\mu_{x}^{2}+\mu_{y}^{2}+(K_{1} L)^{2}\right) \left(\sigma_{x}^{2}+\sigma_{y}^{2}+(K_{2} L)^{2}\right)}},
\end{equation}
where $\mu_{x}$, $\mu_{y}$ and $\sigma_{x}$, $\sigma_{x}$ are the mean and standard deviation and $\sigma_{xy}$ the covariance of all pixel values of two images. To avoid instability, $(K_{1} L)^{2}$ and $(K_{2} L)^{2}$ are added, where $L$ is the dynamic range of the pixel values and $K_{1} = 0.01$ and $K_{2}=0.03$ are small constants. 
SSIM is computed as the average result of a moving $11 \times 11$ kernel with a Gaussian weighting function. The parameters were chosen as suggested by Wang et al.~\cite{wang2004image}.

\textbf{MI}\newcommand{\mi}{MI}: We use Mutual Information (MI) as the second similarity-based approach from information theory. MI is a widely used technique for similarity calculation and registration of medical images~\citep{pluim2003mutual,russakoff2004image} and measures the dependence between two images ${\bf x}$ and ${\bf y}$ by calculating the Kullback–Leibler divergence
\begin{equation}
{\rm KL}({\bf X}\Vert{\bf Y})=\sum_{x\in{\bf X}}\sum_{y\in{\bf Y}}p(x,y)\log \left( \frac{p(x,y)}{p(x) p(y)} \right),
\end{equation}
between the joint distribution $p(x,y)$ and the independent distributions $p(x) p(y)$ of the pixel values. We use the normalized Mutual Information as introduced by Studholme et al.~\citep{studholme1998normalized}.

\textbf{DeepNet}\newcommand{\deep}{DeepNet}: Motivated by the success of the Perceptual Similarity Metric~\citep{zhang2018unreasonable} for image comparison, which is computationally expensive, we introduce \deep{} similarity, which reduces the complexity so that pairwise comparisons can be performed in a reasonable amount of time. Like perceptual similarity, \deep{} similarity compares two images by running them through a pre-trained deep learning model. Instead of computing the cosine similarity in the channel dimension, \deep{} similarity computes the cosine similarity between the output vectors. Using PyTorch's ResNet50~\citep{he2016deep} pre-trained on ImageNet~\citep{deng2009imagenet} to compute the output vectors, we get the following equation  
\begin{equation}
{\rm d}({\bf x},{\bf y}) = \frac{\textrm{ResNet}({\bf x}) \cdot \textrm{ResNet}({\bf y})} {\left\| \textrm{ResNet}({\bf x})\right\| _{2}\left\| \textrm{ResNet}({\bf y})\right\| _{2}},
\end{equation}
to compare two images ${\bf x}$ and ${\bf y}$. 

\textbf{HASH}\newcommand{\hash}{HASH}:
The \hash{} similarity is based on the comparison of hash values derived from each image. It is motivated by Conrad and Narayan~\citep{conrad2021cem500k}, who used it to extract dissimilar images from an electron microscopy dataset for contrastive pre-training. The procedure is as follows: Each image is compressed to the size of $9\times8$ pixel and encoded into a 64-bit hash. The compression is performed with the Antialias function from Pillow~\citep{clark2015pillow}. The hash is computed by looping through each row of the compressed image, comparing each pixel with its right neighbor, and selecting one if the neighbor is larger and zero if the neighbor is smaller. For each row of nine pixels, this results in a hash of eight bits, leading to a 64-bit hash in total. The Hamming distance
\begin{equation}
{\rm d}({\bf x},{\bf y}) = \frac{1}{n}\sum_{i=1}^{n}|\textrm{Hash}({\bf x})_i-\textrm{Hash}({\bf y})_i|,
\end{equation}
between the two hashes of images {\bf x} and {\bf y} measures the similarity, where $n=64$ is the length of the hash. All parameters are chosen following Conrad and Narayan~\citep{conrad2021cem500k}. Figure~\ref{fig:hash} illustrates the similarity calculation between two images with the \hash{} method.

\begin{figure}[!ht]
\centering
\includegraphics[width=0.48\textwidth]{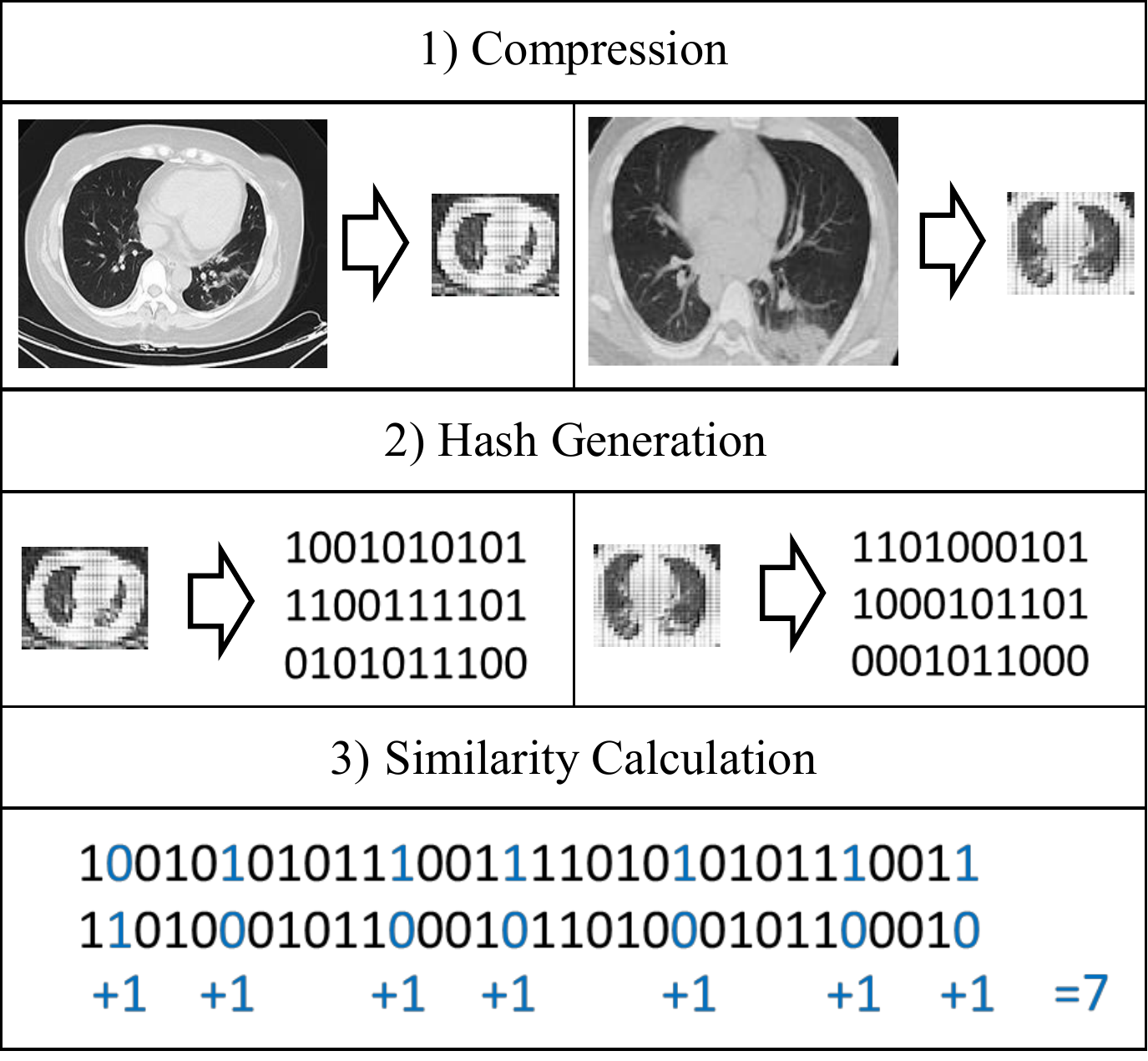}
\caption{This figure explains the similarity calculation between two images using the \hash{} method. First, the images are compressed to the size size of $9\times8$. In the second step, a 64-bit hash is computed by looping through each row of the compressed images, comparing each pixel with its right neighbor, and choosing one if the neighbor is larger and zero if the neighbor is smaller. To calculate the similarity, the hashes for the two images are compared with the Hamming distance, which is the sum of the different bits. }\label{fig:hash}
\end{figure}

\subsection{Pre-Training} \label{sec:PreTrain}
Following Huang et al.`s.~\citep{huang2023self} study, popular contrastive learning methods from natural image processing that are widely utilized for medical pre-training are SimCLR~\citep{chen2020simple}, MoCo~\citep{he2020momentum}, BYOL~\citep{grill2020bootstrap}, and SwAV~\citep{caron2020unsupervised}. SimCLR follows the basic contrastive learning strategy, where the model is trained to distinguish between positive and negative pairs within a mini-batch. This requires a large batch size in order to obtain a sufficient number of negative samples within one mini-batch. MoCo adds a queue for storing negative samples, which reduces batch size requirements but increases storage demands. BYOL introduces two competing models to decrease batch size demands. SwAV adds online feature clustering to the latent space representations, which lowers the batch-size constraints. 

For our evaluations, we chose the method SwAV~\citep{caron2020unsupervised}, since it outperformed the other state-of-the-art contrastive learning methods with convolutional models on several natural imaging benchmark tasks~\citep{tian2023designing}, is more computational efficient~\citep{caron2020unsupervised}, and was already successfully applied for pre-training on CT slices~\citep{wolf2023self,ghesu2022contrastive}. Due to the identical basic concept of all methods, our findings are expected to be generalizable to other contrastive pre-training methods. A detailed explanation of the SwAV pre-training method can be found in~\ref{sec:Appendix1}. In order to create positive pairs of one original image, SwAV uses the transforms color jitter, gaussian blur, and a multi-crop strategy, where two transformed images are obtained by cropping a part of the original image with a larger crop size, and several additional samples are cropped with a smaller crop size. For our evaluations, we use exactly the transform settings of the original paper, as they have been shown to be the most appropriate for this pre-training method. Details can be found in~\ref{sec:Appendix1}.

The pre-training is performed on the CT slices of two publicly available image datasets, summarized in Table~\ref{tab:PreTrain}:

\newcommand{\pet}{PET-CT}\textbf{\pet{}}: The FDG-PET-CT~\citep{gatidis2022data,gatidis2022whole} dataset, which was part of the MICCAI 2022 AutoPET challenge~\citep{sergios_gatidis_2022_6362493}, consists of whole-body PET/CT scans of 900 patients, from which we extract 541,439 CT slices.

\newcommand{\lidc}{LIDC}\textbf{\lidc{}}: The Lung Image Database Consortium Image Collection (LIDC-IDRI)~\citep{armato2011lung,armato2015lung} dataset consists of lung CT volumes of 1,010 patients acquired from seven academic centers, initiated by the National Cancer Institute (NCI).  We extracted 244,527 CT slices of the dataset.

\begin{table}[t]
    \caption{Pre-Training Datasets}
    \centering
    \begin{tabular}{lcc}
                 \toprule             
                     & \pet{} & \lidc{}     \\ 
                 \midrule   
        Modality     & CT         & CT      \\
        Body Part    & Whole Body & Lung    \\
        Volumes      & 900        & 1,010   \\
        Slices       & 541,439    & 244,527 \\
        Availability & Public     & Public  \\
                 \bottomrule    
    \end{tabular}
    \label{tab:PreTrain}
\end{table}

We used only the CT slices of the datasets; all other information or labels were excluded. The pre-training is performed separately on the two datasets for better generalizability of findings. Implementations are done in PyTorch Lightning~\citep{william_falcon_2020_3828935}. We choose a ResNet50~\citep{he2016deep} as our model due to its popularity in medical image analysis~\citep{kora2022transfer} and its widespread use as a baseline for comparisons in vision studies~\citep{liu2022convnet}. We pre-train the model for 800 epochs on an Nvidia GeForce RTX 3090 GPU and perform a downstream task evaluation every 50 epoch to find the best-performing epoch. All pre-training hyperparameters can be found in~\ref{sec:Appendix1}.

\subsection{Downstream Evaluation}

As Huang et al.~\citep{huang2023self} shows, classification tasks are commonly used as a benchmark for evaluating self-supervised pre-training. Usually, only a single linear layer is added to the pre-trained encoder to adjust the model to the correct output size, resulting in only the weights of one layer not being pre-trained. In contrast, segmentation tasks require the addition of a large decoder to the pre-trained encoder, such as in a U-Net~\citep{ronneberger2015u}, resulting in a more significant proportion of untrained model weights. This increases the dependency on the dataset of the downstream task. Therefore, we focus on classification downstream tasks to evaluate pre-training performance, although our results are expected to apply to other tasks as well.

We selected three classification tasks on CT slices, ensuring that the images do not overlap with those in the pre-training datasets. These tasks include two public challenges and an internal task as part of a clinical study.
For the two publicly available challenges, we perform five downstream runs with the given
train/validation/test split of the challenge, to ensure the comparability with other challenge participants.
For the internal task, a five-fold stratified cross-validation is performed. For each fold, four parts of the data are used for training and validation (90\,\% training, 10\,\% validation), and the remaining part that has not been used for training and validation is used for testing. This ensures, that the model works on different data splits. 
The mean and standard deviation of accuracy, AUC score, and F1-sore over the five runs are reported for all three tasks.
The tasks include CT scans from different hospitals, scanners, and body parts to prove the generalizability of our findings. 
The three tasks, summarized in Table~\ref{tab:Downstream}, are the following:

\newcommand{\covid}{COVID-19}\textbf{\covid{}}: 
The COVID-19 CT Classification Grand Challenge~\citep{yang2020covid} dataset consists of 349 CT slices (216 patients) and 397 CT slices (171 patients) with and without clinical findings of COVID-19, respectively. The task is to classify between COVID-19 findings and no COVID-19 findings. Figure~\ref{fig:covid} shows an example slice for both classes. 

\begin{figure}[!ht]
\centering
\includegraphics[width=0.49\textwidth]{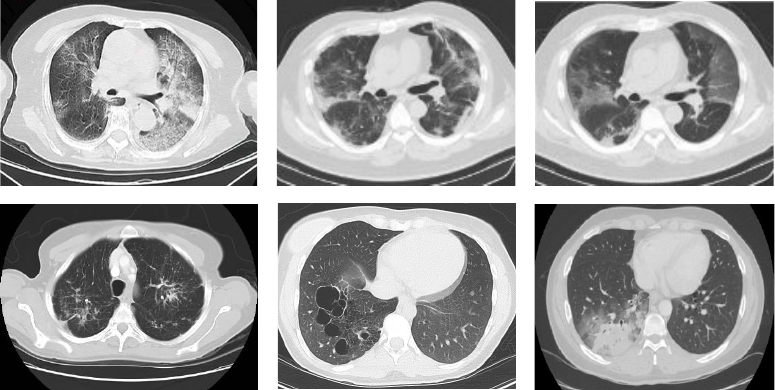}
\caption{Example slices of the \covid{} classification downstream task from Grand Challenge~\citep{yang2020covid}. Upper Row: COVID-19 findings; Lower Row: No COVID-19 findings}\label{fig:covid}
\end{figure}

\newcommand{\mnist}{OrgMNIST}\textbf{\mnist{}}: The OrganSMNIST Challenge from MedMNIST~\citep{medmnistv2} consists of 25,221 image patches of the size $28\times28$, cropped around organs from abdominal CT scans of 201 patients. The challenge is a multi-class classification of 11 body organs. Figure~\ref{fig:mnist} shows some example images of cropped patches. 

\begin{figure}[!ht]
\centering
\includegraphics[width=0.49\textwidth]{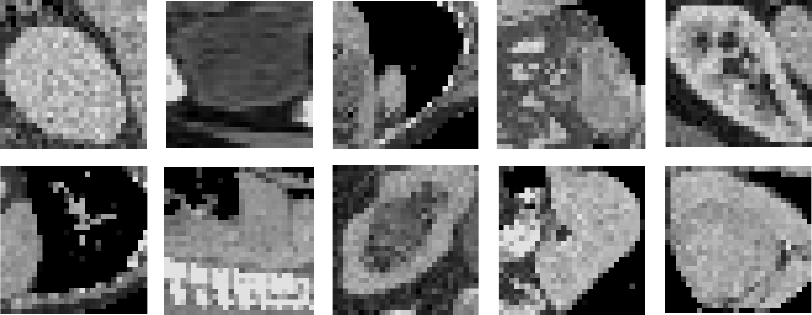}
\caption{Example patches for the \mnist{} multi-class classification downstream task of the OrganSMNIST Challenges~\citep{medmnistv2}}\label{fig:mnist}
\end{figure}

\newcommand{\brain}{Brain}\textbf{\brain{}}:
An internal dataset with CT slices from 100 patients with and 100 patients without brain hemorrhage is used for the third downstream task. All CT examinations were part of the routine clinical practice at the University Hospital of Ulm. Representative slices were selected by Dr. Ch. G. Lisson and Dr. Ca. S. Lisson, two well-trained senior radiologists. This study aims to determine whether brain hemorrhages can be detected automatically on CT scans, which could help physicians in their diagnosis. All patients provided written consent for the use of their anonymized data for research purposes upon signing the treatment contract between the University Hospital of Ulm and the patient. Ethical approval was given by the Ethics Committee of Ulm University under ID 302/17. More details about the collected slices can be found in~\ref{sec:Appendix4}. The task is to classify between brain hemorrhage and no brain hemorrhage, with pre-training being essential due to the small dataset size. Figure~\ref{fig:brain} shows some example images of cropped patches.

\begin{figure}[!ht]
\centering
\includegraphics[width=0.49\textwidth]{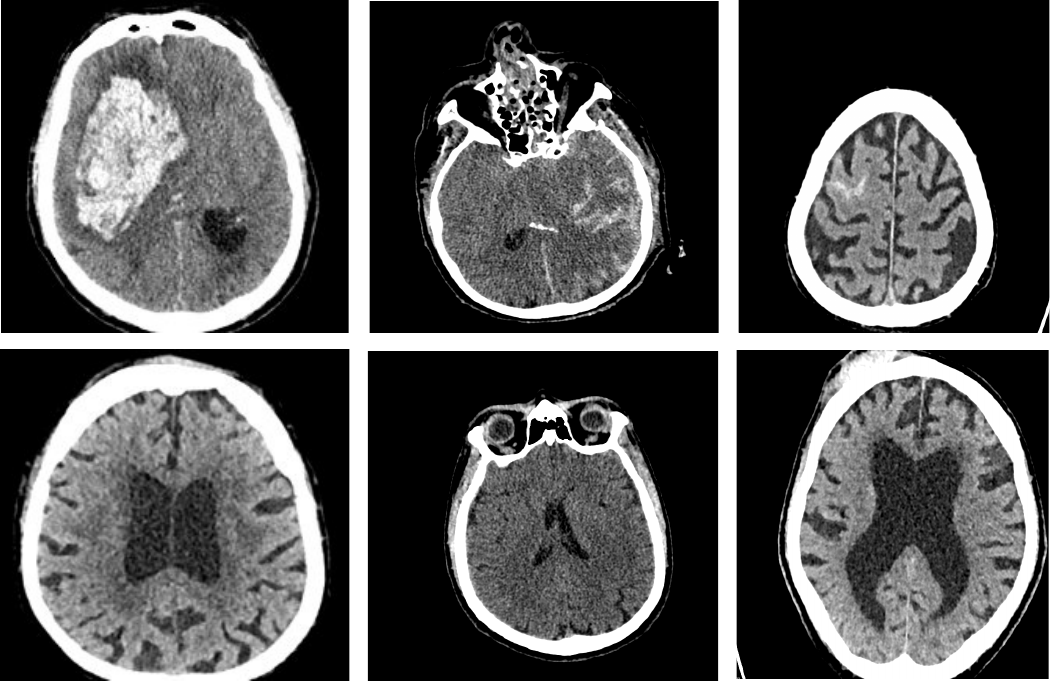}
\caption{Example slices of the internal \brain{} classification downstream task. Upper Row: With brain hemorrhage; Lower Row: Without brain hemorrhage }\label{fig:brain}
\end{figure}

\begin{table}[t]
    \caption{Downstream Tasks to evaluate the Pre-Trainings}
    \centering
    \begin{tabular}{lccc}
                 \toprule             
                     & \covid{}      & \mnist{}    & \brain{}     \\ 
                 \midrule   
        Modality     & CT            & CT          & CT     \\
        Body Part    & Lung          & Abdomen     & Brain   \\
        Classification         & Binary        & Multi-Class & Binary \\
        
        Slices       & 746           & 25,221     & 200 \\
        Availability & Public        & Public     & Internal \\
                 \bottomrule    
    \end{tabular}
    \label{tab:Downstream}
\end{table}

Our implementations are done in PyTorch Lightning~\citep{william_falcon_2020_3828935} with MONAI~\citep{monai_consortium_2022_7086266}. We resize the slices of all tasks to $224\times224$ in a preprocessing step and train on an Nvidia GeForce RTX 3090 GPU using the Adam optimizer with learning rate $10^{-4}$ and batch-size 64. We add one linear layer to the pre-trained encoder. Only the linear layer is trained during the first ten epochs before the complete model is fine-tuned.

\section{Experiments and Results} 

Our experiments are designed to investigate our hypothesis by answering whether dataset reduction leads to performance gains, which of our selected reduction methods performs best, what is the optimal similarity threshold, and how much performance gain can be achieved. All experiments are performed separately on the two pre-training datasets \pet{} and \lidc{}. In total, we conducted 24 different pre-trainings, resulting in over 2000 pre-training hours. The pre-trainings are evaluated on the three downstream tasks \covid{}, \mnist{}, and \brain{}. For all results, we report the mean and standard deviation of AUC and F1 scores over five fine-tuning runs on the downstream tasks. Table~\ref{tab:NoPre} shows the downstream task results without any pre-training as a reference. 

\begingroup
\setlength{\tabcolsep}{5pt}
\begin{table*}[t]
    \caption{This table shows the results of the three downstream tasks \covid{}, \mnist{}, and \brain{} without using any pre-training. The weights of the model are initialized with PyTorch's standard random initialization. (Accuracy can be found in in Table~\ref{tab:NoPreApp} in~\ref{sec:Appendix5})}
    \centering
    \resizebox{\textwidth}{!}{
    \begin{tabular}{l@{\hskip 4pt}l@{\hskip 7pt}cc@{\hskip 7pt}cc@{\hskip 7pt}cc}
        \toprule
         \multicolumn{2}{c}{Pre-Training} & \multicolumn{6}{c}{Downstream Results}  \\
                 \cmidrule(lr){1-2} \cmidrule(lr){3-8}             
        \multicolumn{1}{c}{Dataset} & \multicolumn{1}{c}{Method} &  \multicolumn{2}{c}{\covid{}}  & \multicolumn{2}{c}{\mnist{}} & \multicolumn{2}{c}{\brain{}}  \\
        \cmidrule(lr){3-4} \cmidrule(lr){5-6} \cmidrule(lr){7-8}
        & &  AUC & F1 & AUC & F1 & AUC & F1 \\ 
        \midrule 
     - &      -                 & 0.737 $\pm$ 0.033  & 0.679 $\pm$ 0.033 &   0.971 $\pm$ 0.001 & 0.755 $\pm$ 0.003 & 0.678 $\pm$ 0.037 & 0.447 $\pm$ 0.157\\
        \bottomrule
    \end{tabular}
    }
    \label{tab:NoPre}
\end{table*}
\endgroup

\subsection{Evaluation A: Does reduction lead to performance gains?} The first experiment evaluates whether reducing CT datasets for contrastive pre-training leads to performance gains in downstream tasks. To answer this question, we compare the baseline method \all{} with the baseline reduction method \everyn{}. Pre-training is performed on both pre-training datasets with all slices (\all{}), with every tenth slice, and with every fifth slice (\everyn{}). 
The reduction numbers are chosen randomly.
Table~\ref{tab:baseline} shows the downstream task results. Performance gains are achieved in all three downstream tasks by reducing the pre-training dataset to 20\,\%, and 10\,\% with the \everyn{} method. The performance gains are slightly higher for the 10\,\% reduction.

\begingroup
\setlength{\tabcolsep}{2.5pt}
\begin{table*}[t]
    \caption{Evaluation A: This table compares the baseline pre-training method \all{}, the current state-of-the-art, which uses all slices of a CT dataset for contrastive pre-training, with the baseline reduction pre-training method \everyn{}.
    Pre-training with SwAV is performed on the datasets \pet{} and \lidc{} with all slices, with 20\,\% of the slices by using every fifth slice, and with 10\,\% of the slices, by using every tenth slice. The different pre-trainings are evaluated on the three downstream tasks \covid{}, \mnist{}, and \brain{}. Accuracy can be found in in Table~\ref{tab:baselineApp} in~\ref{sec:Appendix5})}
    \centering
    \resizebox{\textwidth}{!}{
    \begin{tabular}{l@{\hskip 4pt}l@{\hskip 7pt}cc@{\hskip 7pt}cc@{\hskip 7pt}cc}
        \toprule
         \multicolumn{2}{c}{Pre-Training} & \multicolumn{6}{c}{Downstream Results}  \\
                 \cmidrule(lr){1-2} \cmidrule(lr){3-8}             
        \multicolumn{1}{c}{Dataset} & \multicolumn{1}{c}{Method} &  \multicolumn{2}{c}{\covid{}}  & \multicolumn{2}{c}{\mnist{}} & \multicolumn{2}{c}{\brain{}}  \\
        \cmidrule(lr){3-4} \cmidrule(lr){5-6} \cmidrule(lr){7-8}
        & &  AUC & F1 & AUC & F1 & AUC & F1 \\ 
        \midrule 
      \pet{}  & \all{}            & 0.775 $\pm$ 0.009 & 0.719 $\pm$ 0.010 & 0.968 $\pm$ 0.003 & 0.752 $\pm$ 0.003 & 0.727 $\pm$ 0.042 & 0.534 $\pm$ 0.073\\
              & \everyn{} 20\,\%  & 0.801 $\pm$ 0.006  & 0.735 $\pm$ 0.009 & 0.972 $\pm$ 0.003  & 0.782 $\pm$ 0.003 & 0.781 $\pm$ 0.035 & 0.665 $\pm$ 0.070 \\
              & \everyn{} 10\,\%  & \textbf{0.810 $\pm$ 0.007}  & \textbf{0.740 $\pm$ 0.016} & \textbf{0.973 $\pm$ 0.002}  & \textbf{0.793 $\pm$ 0.002} & \textbf{0.798 $\pm$ 0.031} & \textbf{0.674 $\pm$ 0.074}\\
      \rule{0pt}{3ex}%
      \lidc{} & \all{}            & 0.807 $\pm$ 0.006 & 0.744 $\pm$ 0.013 & 0.972 $\pm$ 0.003 & 0.769 $\pm$ 0.003 & 0.734 $\pm$ 0.046 & 0.609 $\pm$ 0.072\\
              & \everyn{} 20\,\%  & 0.810 $\pm$ 0.004 & 0.751 $\pm$ 0.010 & 0.977 $\pm$ 0.005 & 0.792 $\pm$ 0.003 & 0.739 $\pm$ 0.044 & 0.610 $\pm$ 0.046\\
              & \everyn{} 10\,\%  & \textbf{0.812 $\pm$ 0.006} & \textbf{0.756 $\pm$ 0.010} & \textbf{0.979 $\pm$ 0.002} & \textbf{0.800 $\pm$ 0.003} & \textbf{0.740 $\pm$ 0.041} & \textbf{0.614 $\pm$ 0.046}\\
        \bottomrule
    \end{tabular}
    }
    \label{tab:baseline}
\end{table*}
\endgroup

\subsection{Evaluation B: Which reduction method performs best?} 
Having found that CT data reduction for contrastive pre-training leads to considerable performance gains in downstream tasks, the second experiment investigates which of our selected reduction methods is the best option. We compare the baseline reduction method \everyn{} with the similarity-based approaches \ssim{}, \mi{}, \deep{}, and \hash{}. For an accurate comparison, the reduced datasets should contain the same number of slices for each method. We chose to reduce the pre-training datasets to 10\,\%, since we found a considerable performance gain for reducing the datasets to 10\,\% with the baseline method. 
The similarity-based approaches reduce the dataset by performing a pairwise comparison of all slices in a volume and removing one slice from pairs with a high similarity, starting with the highest similarity until only 10\,\% of the volume is left. This results in ten pre-training datasets, the reduced \pet{} and \lidc{} datasets with the approaches \everyn{}, \ssim{}, \mi{}, \deep{}, \hash{}.

\begin{figure*}[!ht]
\centering
\includegraphics[width=1\textwidth]{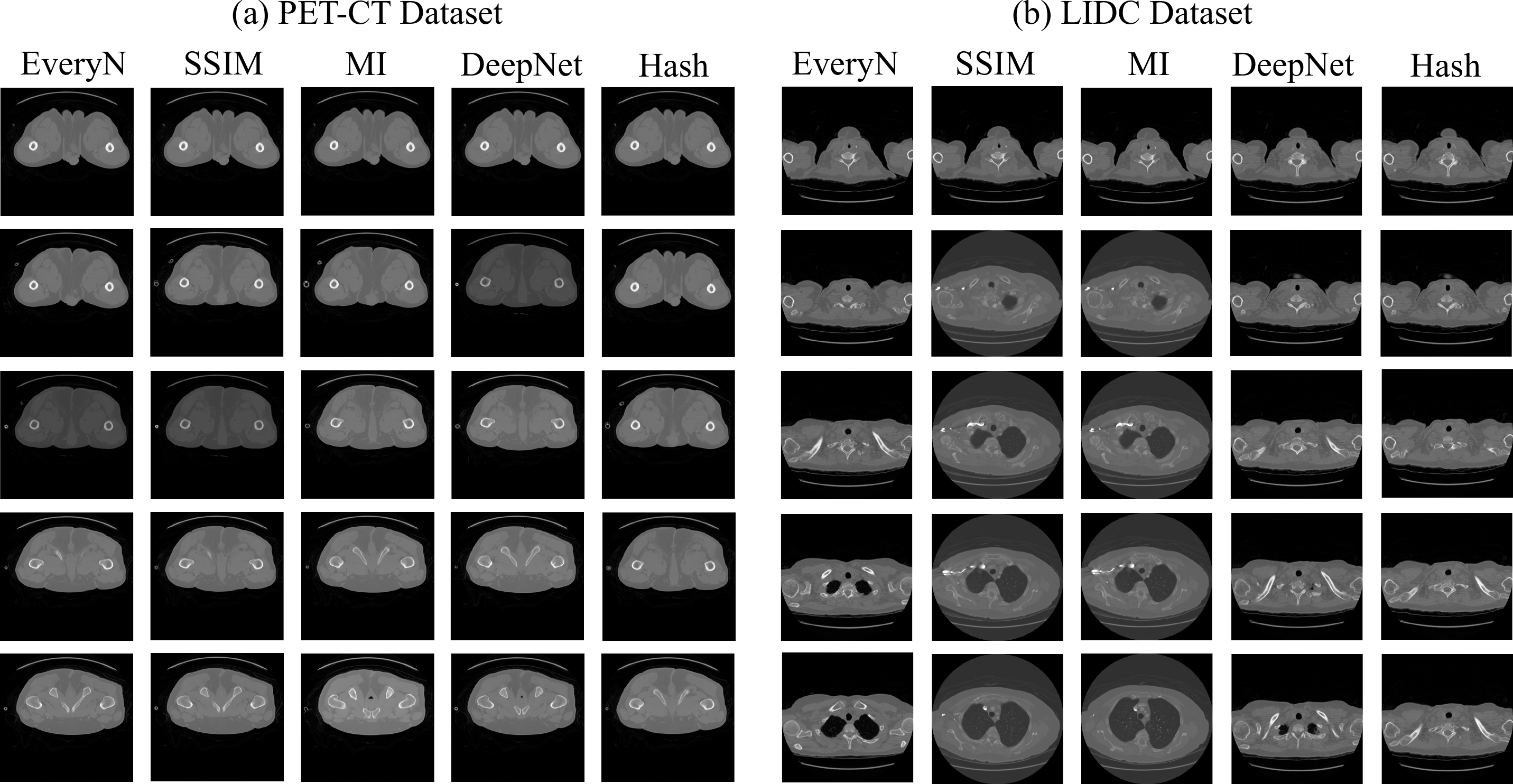}
\caption{This figure compares the selected images for pre-training with the \pet{} dataset (a) and the \lidc{} dataset (b) after the reduction using the \everyn{}, \ssim{}, \mi{}, \deep{}, \hash{} methods. Five consecutive slices from an example volume are shown, starting from the first slice of the volume in the top row to the fifth remaining slice in the bottom row.}\label{fig:Compare}
\end{figure*}

Table~\ref{tab:methods} shows the downstream task results. The \hash{} method outperforms the baseline reduction method \everyn{} and all other similarity based approaches.  We examined the remaining slices after reduction. Figure~\ref{fig:Compare} shows the first five remaining slices for an example volume for each of the two pre-training datasets \pet{} and \lidc{}. To examine how alike the datasets are after the different reduction methods, we compare each dataset with all other datasets and count how many of the remaining slices are equal. The percentage of equal slices across the reduction methods ranges from 9\% to 30\%, with SSIM and MI having the highest equality and the equality between EveryN and the other approaches being the lowest, between 9\% and 11\%.
We further evaluated the execution time for dataset reduction. The \everyn{} approach has the shortest execution time with less than one minute, followed by \hash{} with less than 30 minutes, both executed on an AMD Ryzen 9 5900X. \ssim{}, \mi{}, and \deep{} are computed on an Nvidia GeForce RTX 3090 GPU with execution times of 421\,h, 312\,h, 6\,h for the \pet{} dataset and 53\,h, 48\,h, 2\,h for the \lidc{} dataset. 

\begingroup
\setlength{\tabcolsep}{2.5pt}
\begin{table*}[t]
    \caption{Evaluation B: This table compares different methods for reducing the pre-training datasets to 10\,\% of the slices. The first method is the baseline reduction method \everyn{}, which reduces the pre-training dataset by using every tenth slice, followed by the similarity based methods, which perform a pairwise comparison of all slices in a CT volume and remove one slice from pairs with high similarity.(Accuracy can be found in in Table~\ref{tab:methodsApp} in~\ref{sec:Appendix5})}
    \centering
    \resizebox{\textwidth}{!}{
    \begin{tabular}{l@{\hskip 4pt}l@{\hskip 7pt}cc@{\hskip 7pt}cc@{\hskip 7pt}cc}
        \toprule
         \multicolumn{2}{c}{Pre-Training} & \multicolumn{6}{c}{Downstream Results}  \\
                 \cmidrule(lr){1-2} \cmidrule(lr){3-8}             
        \multicolumn{1}{c}{Dataset} & \multicolumn{1}{c}{Method} &  \multicolumn{2}{c}{\covid{}}  & \multicolumn{2}{c}{\mnist{}} & \multicolumn{2}{c}{\brain{}}  \\
        \cmidrule(lr){3-4} \cmidrule(lr){5-6} \cmidrule(lr){7-8}
        & &  AUC & F1 & AUC & F1 & AUC & F1 \\
        \midrule
      \pet{}   & \everyn{}   & 0.810 $\pm$ 0.007 &  0.740 $\pm$ 0.016 & 0.973 $\pm$ 0.002 & 0.793 $\pm$ 0.002 & 0.798 $\pm$ 0.031 & 0.674 $\pm$ 0.074\\
               & \ssim{}     & 0.811 $\pm$ 0.005 & 0.741 $\pm$ 0.010 & 0.974 $\pm$ 0.001 & 0.794 $\pm$ 0.002 & 0.801 $\pm$ 0.309 & 0.701 $\pm$ 0.309\\
               & \mi{}       & 0.810 $\pm$ 0.006 & 0.748 $\pm$ 0.005 & 0.974 $\pm$ 0.001 & 0.794 $\pm$ 0.002 & 0.819 $\pm$ 0.020 & 0.720 $\pm$ 0.024\\
               & \deep{}     & 0.791 $\pm$ 0.008 & 0.734 $\pm$ 0.005 & 0.973 $\pm$ 0.002 & 0.795 $\pm$ 0.002 & 0.814 $\pm$ 0.020 & 0.721 $\pm$ 0.011\\
               & \hash{}     & \textbf{0.825 $\pm$ 0.004} & \textbf{0.755 $\pm$ 0.009} & \textbf{0.975 $\pm$ 0.001} & \textbf{0.800 $\pm$ 0.002} & \textbf{0.821 $\pm$ 0.009} & \textbf{0.725 $\pm$ 0.009}\\
          \rule{0pt}{3ex}%
      \lidc{} & \everyn{}   & 0.812 $\pm$ 0.006 & \textbf{0.756 $\pm$ 0.010} & 0.979 $\pm$ 0.002 & 0.800 $\pm$ 0.003 & 0.740 $\pm$ 0.041 & 0.614 $\pm$ 0.046\\
              & \ssim{}     & 0.820 $\pm$ 0.005 & 0.751 $\pm$ 0.008 & 0.980 $\pm$ 0.001 & 0.799 $\pm$ 0.003 & 0.813 $\pm$ 0.031 & 0.740 $\pm$ 0.027\\
              & \mi{}       & 0.820 $\pm$ 0.007 & 0.752 $\pm$ 0.010 & 0.980 $\pm$ 0.001 & 0.800 $\pm$ 0.002 & 0.803 $\pm$ 0.021 & 0.741 $\pm$ 0.025\\
              & \deep{}     & 0.800 $\pm$ 0.005 & 0.744 $\pm$ 0.011 & 0.978 $\pm$ 0.002 & 0.793 $\pm$ 0.002 & 0.817 $\pm$ 0.028 &0.742 $\pm$ 0.064\\
              & \hash{}     & \textbf{0.825 $\pm$ 0.007} & 0.754 $\pm$ 0.013 & \textbf{0.981 $\pm$ 0.002} & \textbf{0.802 $\pm$ 0.002} & \textbf{0.829 $\pm$ 0.020} & \textbf{0.744 $\pm$ 0.021} \\
        \bottomrule
    \end{tabular}
    }
    \label{tab:methods}
\end{table*}
\endgroup

\subsection{Evaluation C: What is the optimal threshold?} When comparing five approaches for reducing CT datasets to 10\,\%, we found that the \hash{} approach performs best. However, the percentage of a CT dataset volume that leads to the best results can vary from dataset to dataset, depending on the variation of the datasets. Datasets with high variation, for example, due to higher slice thickness, may require less reduction than datasets with lower variation. Therefore, in the third experiment, we attempt to find the optimal threshold for the degree of similarity between the slices that leads to the highest results in downstream tasks. We use the best-performing slice selection method \hash{} and test different similarity thresholds. The similarity score for comparing two slices using the \hash{} approach is the Hamming distance, which ranges from 0 (most similar) to 64 (most dissimilar). Reducing the dataset to a chosen similarity threshold of the Hamming distance leads to a dataset where, within each volume, no pairs of slices are more similar than the threshold. We compare three thresholds: Hamming distances three, six, and twelve. Number of slices: \pet{}: 120,750 (22.3\%); 48,718 (9\%); 19,497 (3.6\%) and \lidc{}: 44,416 (18.2\%); 22,672 (9.3\%); 9,828 (4\%).
Table~\ref{tab:threshold} compares the downstream results of the different similarity thresholds. The best performance on all three downstream tasks is achieved with threshold hamming distance six. Higher and lower similarity thresholds, resulting in larger and smaller remaining portions of the pre-training datasets, lead to slightly degraded results. 

\begingroup
\setlength{\tabcolsep}{2.5pt}
\begin{table*}[t]
    \caption{Evaluation C: This table compares different similarity thresholds of the best performing reduction method \hash{}, in order to obtain the optimal degree of similarity between the slices for contrastive pre-training. (Accuracy can be found in in Table~\ref{tab:thresholdApp} in~\ref{sec:Appendix5})}
    \centering
\resizebox{\textwidth}{!}{
    \begin{tabular}{l@{\hskip 4pt}l@{\hskip 7pt}cc@{\hskip 7pt}cc@{\hskip 7pt}cc}
        \toprule
         \multicolumn{2}{c}{Pre-Training} & \multicolumn{6}{c}{Downstream Results}  \\
                 \cmidrule(lr){1-2} \cmidrule(lr){3-8}             
        \multicolumn{1}{c}{Dataset} & \multicolumn{1}{c}{Method} &  \multicolumn{2}{c}{\covid{}}  & \multicolumn{2}{c}{\mnist{}} & \multicolumn{2}{c}{\brain{}}  \\
        \cmidrule(lr){3-4} \cmidrule(lr){5-6} \cmidrule(lr){7-8}
        & &  AUC & F1 & AUC & F1 & AUC & F1 \\
        \midrule 
      \pet{}  &  \hash{} - 3   & 0.821 $\pm$ 0.004  & 0.764 $\pm$ 0.004 & 0.976 $\pm$ 0.001  & 0.797 $\pm$ 0.003 & 0.799 $\pm$ 0.012 & 0.687 $\pm$ 0.018\\
              &  \hash{} - 6   & \textbf{0.830 $\pm$ 0.006}  & \textbf{0.777 $\pm$ 0.016} & \textbf{0.978 $\pm$ 0.001} & \textbf{0.800 $\pm$ 0.003} & \textbf{0.831 $\pm$ 0.021} & \textbf{0.765 $\pm$ 0.027} \\
              &  \hash{} - 12  & 0.822 $\pm$ 0.003  & 0.755 $\pm$ 0.006 & 0.976 $\pm$ 0.001  & 0.796 $\pm$ 0.002 & 0.770 $\pm$ 0.030 & 0.697 $\pm$ 0.039\\
             \rule{0pt}{3ex}%
      \lidc{} &  \hash{} - 3   & 0.813 $\pm$ 0.004  & 0.730 $\pm$ 0.009 & 0.981 $\pm$ 0.001 & 0.800 $\pm$ 0.002 & 0.790 $\pm$ 0.008 & 0.723 $\pm$ 0.041 \\
              &  \hash{} - 6   & \textbf{0.823 $\pm$ 0.005}  & \textbf{0.768 $\pm$ 0.008} & \textbf{0.982 $\pm$ 0.001} & \textbf{0.802 $\pm$ 0.002} & \textbf{0.840 $\pm$ 0.016} & \textbf{0.800 $\pm$ 0.033} \\
              &  \hash{} - 12  & 0.811 $\pm$ 0.003  & 0.737 $\pm$ 0.013 & 0.980 $\pm$ 0.001 & 0.798 $\pm$ 0.002 & 0.798 $\pm$ 0.026 & 0.677 $\pm$ 0.026 \\

        \bottomrule
    \end{tabular}
    }
    \label{tab:threshold}
\end{table*}
\endgroup 

\subsection{Evaluation D: How much performance gain can be achieved?} 
\begin{figure*}[!ht]
\centering
\includegraphics[width=1\textwidth]{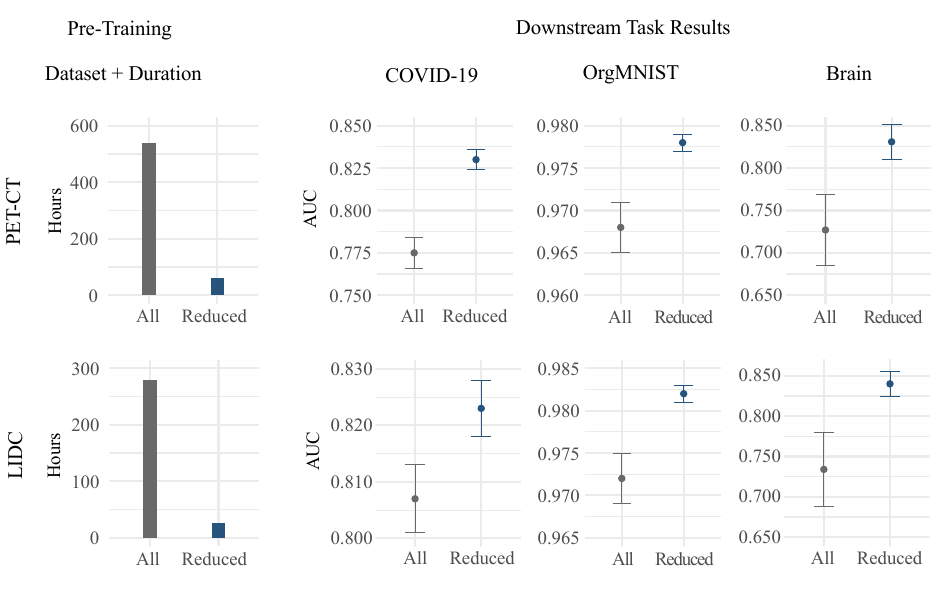}
\caption{Here, we show on two pre-training CT datasets (PET-CT, LIDC) and three downstream CT classification tasks to evaluate the pre-trainings (COVID-19, OrgMNIST, Brain) that our proposed hashing based dataset reduction method leads to shorter pre-training duration and improves downstream task performances, compared to the current state-of-the-art approach, which uses pre-training with all slices of the dataset.}\label{fig:results} 
\end{figure*}

Through several experiments, we found that the \hash{} approach with a Hamming distance threshold of six (\hash{}-6) performs best. In the last step, we compare the downstream task results of the best-performing approach with the baseline method \all{}, the current state of the art, using all slices of the dataset for pre-training. 
Figure~\ref{fig:results} shows the pre-training duration and the AUC scores of the downstream task results. 
On the \pet{} pre-training dataset, we achieve performance gains in AUC values from 0.775 to 0.830, 0.968 to 0.978, and 0.727 to 0.831 for the \covid{}, \mnist{}, and \brain{} downstream tasks, respectively. Performance gains from 0.807 to 0.823, 0.972 to 0.982, and 0.734 to 0.840 are achieved on the \lidc{} pre-training dataset. The pre-training time is reduced from 538\,h to 62\,h and from 280\,h to 27\,h on the \pet{} and \lidc{} datasets, respectively, with a slice selection time of less than 30 minutes. For a better interpretation of our results, in the following we further analyze the difference between pre-training with \all{} data and pre-training with the best performing method \hash{}-6 by calculating the Centered Kernel Alignment CKA~\citep{kornblith2019similarity}, visualizing the t-Distributed Stochastic Neighbor Embedding~\citep{van2008visualizing} and visualizing the model`s attentions with Grad-Cam~\citep{selvaraju2017grad}.

\begin{figure}[!ht]
\centering
\includegraphics[width=0.49\textwidth]{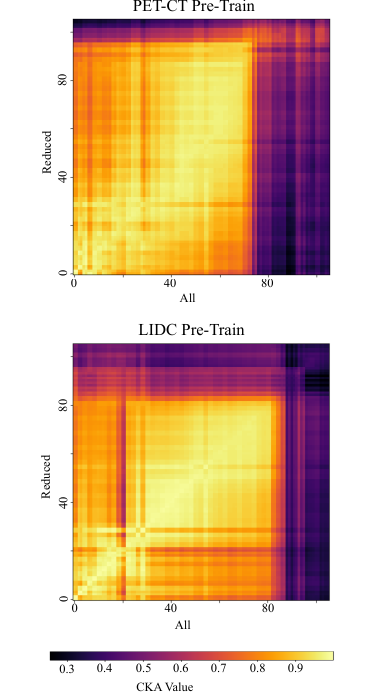}
\caption{Here we visualize the Centered Kernel Alignment CKA~\citep{kornblith2019similarity} between the model pre-trained with \all{} data and the model pre-trained with \hash{}-6 reduced data, for both pre-training datasets \pet{} and \lidc{}. The calculations are done after pre-training, before fine-tuning the model. On the x- and y-axis are the layers of the models starting from zero as the first layer up to the last layer of the model. At the bottom is a scale of the CKA value. High values of the CKA mean that the representations of the two models are similar. The calculations are done by CKA.pytorch~\citep{kim2023stability}.}\label{fig:CKA_pre_pre}
\end{figure}

\begin{figure*}[!ht]
\centering
\includegraphics[width=1\textwidth]{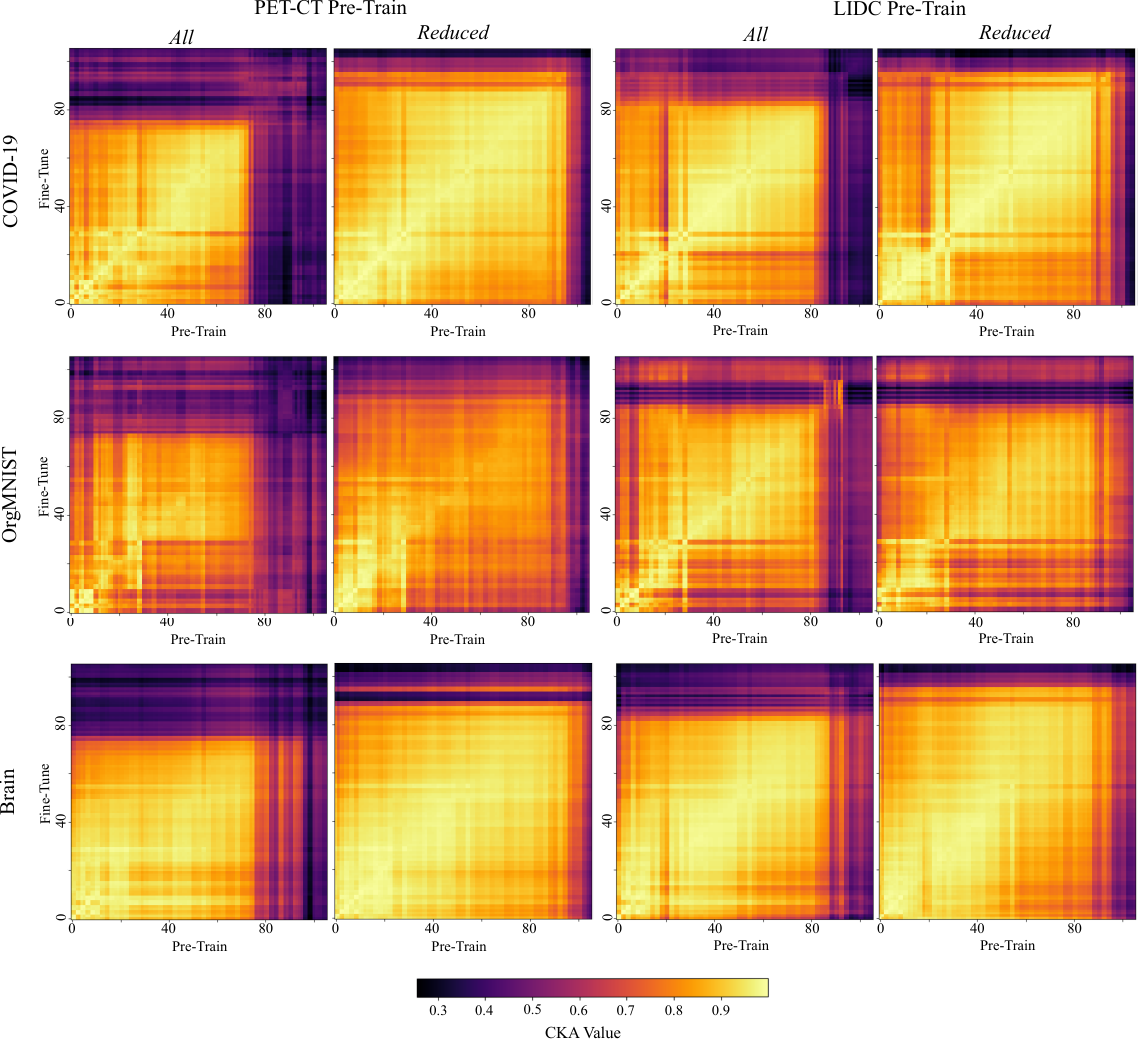}
\caption{Here we visualize the Centered Kernel Alignment CKA~\citep{kornblith2019similarity} between the model after pre-training and the model after fine-tuning. We show the plots for fine-tuning on the three downstream tasks \covid{}, \mnist{}, and \brain{} and the two pre-training datasets \pet{} and \lidc{}, each once for pre-training with \all{} data and once for pre-training with the \hash{}-6 reduced data. At the bottom is a scale of the CKA value. High values of the CKA mean that the representations of the two models are similar. The calculations are done by CKA.pytorch~\citep{kim2023stability}.}\label{fig:CKA_pre_fine}
\end{figure*}

Centered Kernel Alignment CKA~\citep{kornblith2019similarity} measures the similarity of the representations from two models at the different layers of the models. In Figure~\ref{fig:CKA_pre_pre} we show the CKA between pre-training with \all{} data and pre-training with \hash{}-6 reduced data. The calculation was done directly after pre-training, before fine-tuning the model for a specific downstream task. For both pre-training datasets \pet{} and \lidc{}, the plots show a relatively high similarity for early layers of the model, but a relatively low similarity for later layers. Thus, for both pre-training datasets, the reduction of the dataset mainly affects the later layers of the model. In Figure~\ref{fig:CKA_pre_fine} we show the CKA similarity between the model after pre-training and the model after fine-tuning. As can be seen in the plots, fine-tuning mainly affects the later layers of the model while the earlier layers keep similar representations to the stage after pre-training. Now we compare in Figure~\ref{fig:CKA_pre_fine} the CKA plots of pre-training with \all{} data to the CKA plots of pre-training with \hash{}-6 reduced data. This comparison shows that with the \hash{}-6 reduced data, the representations of more layers have a high similarity between pre-training and fine-tuning, compared to pre-training with \all{} data. So the Centered Kernel Alignment CKA between pre-training and fine-tuning is higher when the \hash{}-6 reduced dataset is used for pre-training.

\begin{figure*}[!ht]
\centering
\includegraphics[width=1\textwidth]{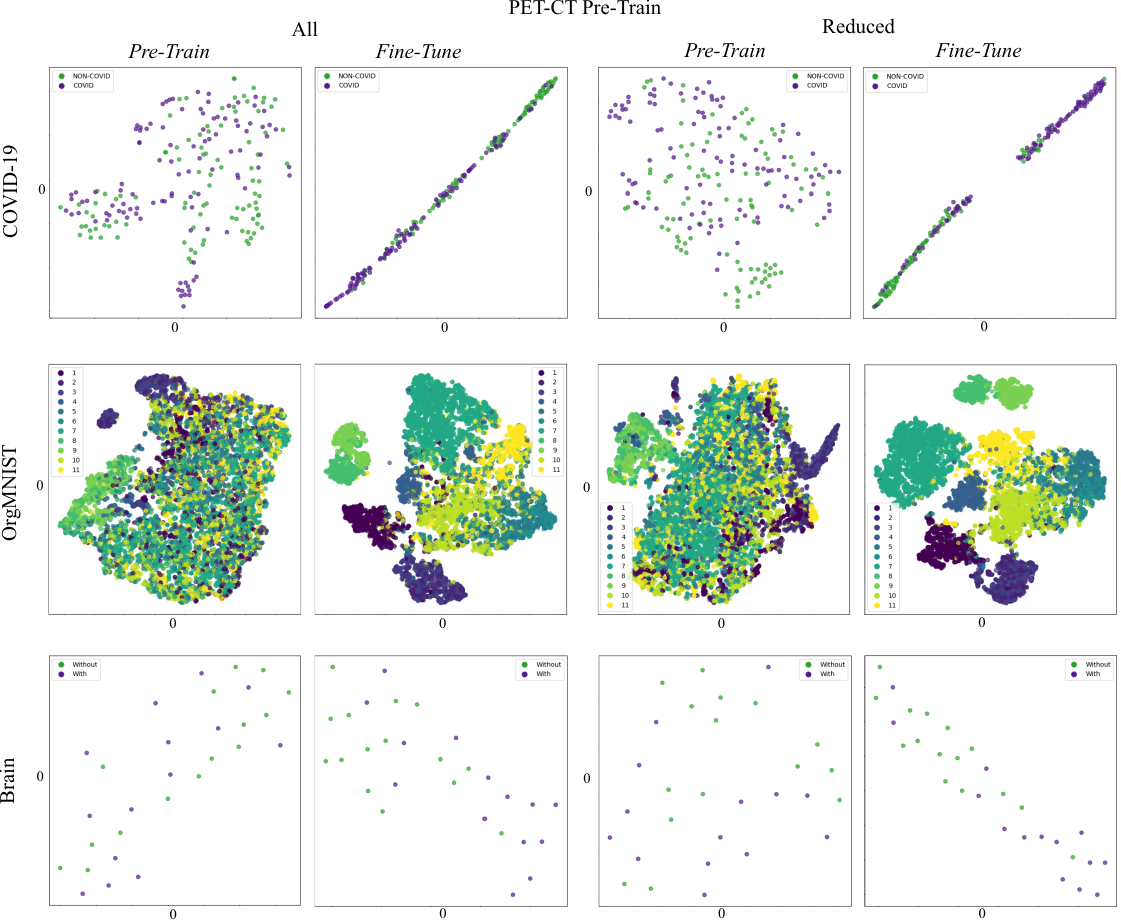}
\caption{Here we visualize the T-Distributed Stochastic Neighbor Embedding (t-SNE)~\citep{van2008visualizing} at the fully connected classifier output after the convolutional layers of our model when the test dataset images of the downstream tasks \covid{}, \mnist{}, \brain{} are propagated through the model. The plots were generated once after pre-training and once after fine-tuning. On the left, the model was pre-trained with \all{} data, and on the right with \hash{}-6 reduced data. The colors of the dots indicate the target classes of the images. \covid{} and \brain{} are binary and \mnist{} is multi-class classification tasks. The calculations are done with scikit-learn ~\citep{pedregosa2011scikit}. }\label{fig:SNE_PET}
\end{figure*}

\begin{figure*}[!ht]
\centering
\includegraphics[width=1\textwidth]{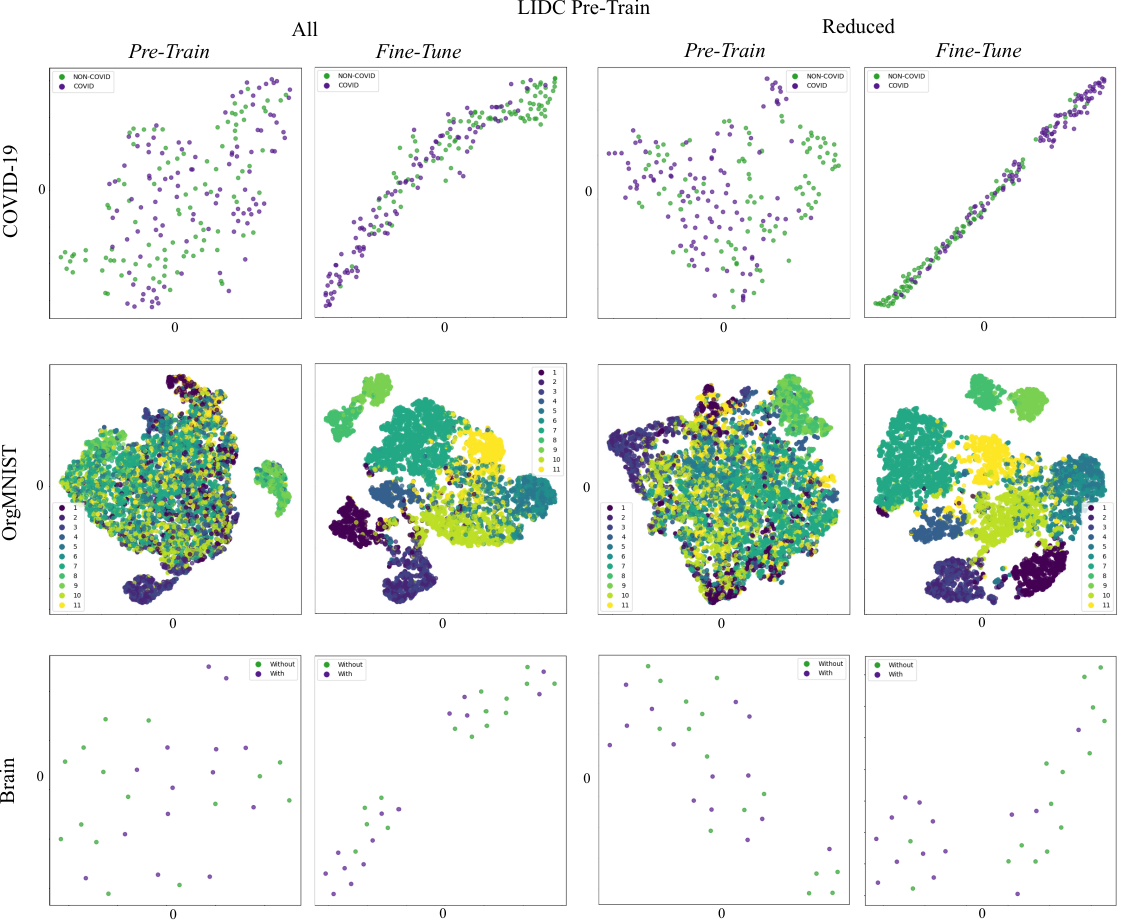}
\caption{Here we visualize the T-Distributed Stochastic Neighbor Embedding (t-SNE)~\citep{van2008visualizing} at the fully connected classifier output after the convolutional layers of our model when the test dataset images of the downstream tasks \covid{}, \mnist{}, \brain{} are propagated through the model. The plots were generated once after pre-training and once after fine-tuning. On the left, the model was pre-trained with \all{} data, and on the right with \hash{}-6 reduced data. The colors of the dots indicate the target classes of the images. \covid{} and \brain{} are binary and \mnist{} is multi-class classification tasks.  The calculations are done with scikit-learn ~\citep{pedregosa2011scikit}. }\label{fig:SNE_LIDC}
\end{figure*}

T-Distributed Stochastic Neighbor Embedding (t-SNE)~\citep{van2008visualizing} is a technique for visualizing high-dimensional data by reducing the data to lower-dimensional spaces. We use this technique to visualize the fully connected classifier output at the end of our model after the convolutional layers. We propagate the images of the test datasets of the three downstream tasks \covid{}, \mnist{}, and \brain{} through the model up to the fully connected layer and plot the values with t-SNE to visualize the distributions of the predictions and see how well the model can discriminate between the classes. This visualization was done once directly after pre-training, before fine-tuning the model, and once after fine-tuning. Figure~\ref{fig:SNE_PET} shows the plots for the \pet{} pre-training dataset and Figure~\ref{fig:SNE_LIDC} for the \lidc{} pre-training dataset. We again compare pre-training with \all{} data and pre-training with \hash{}-6 reduced data. After pre-training, there is no clear separation of the different classes, neither for pre-training with \all{} data nor for pre-training with the \hash{}-6 reduced data. After fine-tuning, especially for the \covid{} and the \mnist{} task, a clearer separation of the classes is visible when using \hash{}-6 reduced pre-training compared to \all{} pre-training. As a quantitative measure, we calculated, for fine-tuning, the Pearson Correlation Coefficient (PCC) between the t-SNE values of the model and the target classes. We get between 2\% and 16\% higher PCC values after fine-tuning when pre-training with the \hash{}-6 reduced data compared to pre-training with \all{} data. Thus, the distribution of predictions indicates that the model pre-trained with \hash{}-6 reduced data can better discriminate between classes after fine-tuning compared to \all{} data pre-training.

\begin{figure*}[!ht]
\centering
\includegraphics[width=1\textwidth]{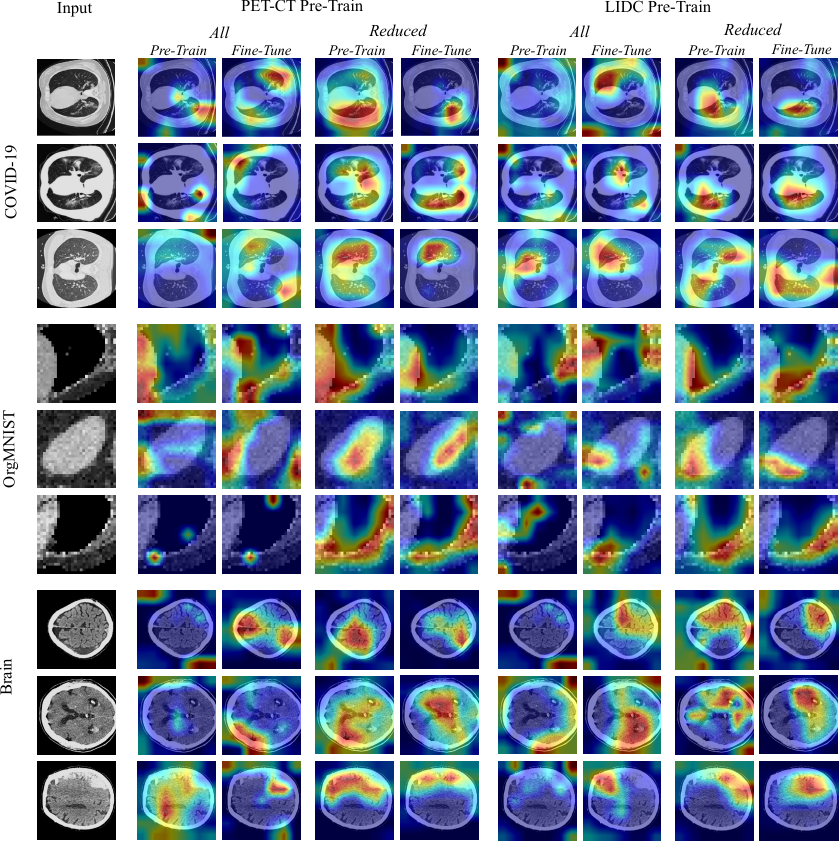}
\caption{Here we visualize the attention region of the network with Grad-Cam~\citep{selvaraju2017grad} for three example images of the \covid{}, \mnist{} and \brain{} downstream tasks` test datasets. The first row shows the input image, the next four rows the attention heatmaps with the \pet{} pre-training dataset, and the last four rows the attention maps with the \lidc{} pre-training dataset. For both pre-training datasets, we visualize the heatmaps for pre-training with \all{} data and for the best reduction approach \hash{}-6. We compare the attention heatmap after the self-supervised pre-training before fine-tuning the model with the attention heatmap after the fine-tuning for the specific downstream task. The plots are generated by pytorch-grad-cam~\citep{jacobgilpytorchcam}. }\label{fig:GradCam}
\end{figure*}

We visualize the attention region of the model with the gradient-weighted class activation mapping Grad-Cam~\citep{selvaraju2017grad}. We generated the attention heatmaps on the test datasets of the three downstream tasks for both pre-training datasets. The attention heatmaps were generated once directly after pre-training, before fine-tuning the model, and once after fine-tuning. Figure~\ref{fig:GradCam} shows three example images for each downstream task. 
The attention regions were qualitatively analyzed by two well-trained radiologists. As can be seen in the example images in Figure~\ref{fig:GradCam} for the \covid{} and \brain{} downstream task, the main attention after pre-training with \all{} data is often not even in the lung or brain region and is far away from the model's final attention after fine-tuning. In contrast, when pre-training with the \hash{}-6 reduced, the model`s main attention is already after pre-training mostly much closer to the actual target and the final attention.
For example, for the \covid{} task, in row 4, column 2, the Grad-cam image for \all{} data after pre-training shows an attention that lies outside the body region and is far away from the final attention after fine-tuning (image row 4, column 3). In contrast, in the Grad-cam image for the \hash{}-6 reduced data after pre-training (row 4, column 4), the attention is clearly in the lung region and already close to the final attention after fine-tuning (row 4, column 5). And the final attention for the \hash{}-6 reduced data covers the area that the well-trained radiologists would look at much better. 
For the \mnist{} task, for example in row 6, column 6, the attention after pre-training with \all{} data is somewhere completely different from the final attention in column 7. Meanwhile, with the \hash{}-6 reduced data, the attention after pre-training is already relatively close to the final attention after fine-tuning (row 6, column 8 and 9). 
For the \brain{} task in the last row, column 2, the attention after pre-training with \all{} data is widely distributed over the image and not close to the final attention after fine-tuning (last row, column 3). And even the final attention does not cover the bleeding perfectly. In contrast, after pre-training with \hash{}-6 reduced data (last row, column 4), the attention is in the brain area and already covers the bleeding almost perfectly. After fine-tuning, the attention becomes only slightly more precise (last row, column 5).
The same pattern can be seen for most of the Grad-cam images on both pre-training datasets and all three downstream tasks.
For a quantitative analysis, we computed the Intersection over Union (IoU) between the heatmap after pre-training and the heatmap after fine-tuning, to see how close the model's attention after pre-training is already to the model's final attention after fine-tuning. Again, we compared pre-training with \all{} data and pre-training with \hash{}-6 reduced data. For all three downstream tasks, on both pre-training datasets, we get between 7\% and 9\% higher IoU values with the reduced pre-training dataset, as with all pre-training data. Thus, with our \hash{}-6 reduction approach, the model's attention after pre-training is already closer to the final attention after fine-tuning.

\subsection{Evaluation E: Other Self-Supervised Pre-Training Approaches}
To show the generalizability of our results, we tested further self-supervised pre-training approaches with the best-performing reduction approach \hash{}. For this evaluation, we only use the LIDC dataset which is smaller and thus needs less pre-training time and has less computational effort. 
According to Huang et al.`s.~\citep{huang2023self} study, another popular contrastive learning approach on convolutional neural networks from natural image processing that is widely utilized for medical pre-training is MoCo (Momentum Contrast)~\citep{he2020momentum}. MoCo has been slightly updated in MoCo version 2~\citep{chen2020improved} and has also been successfully applied to pre-training on CT slices~\citep{wolf2023self,chen2021momentum}.   
A completely different approach for self-supervised pre-training is masked image modeling, which has gained much popularity in the imaging field in the last few years~\citep{he2022masked}.
In a recent study, Tian et al.~\citep{tian2023designing} show on ImageNet~\citep{deng2009imagenet} data that masked autoencoders~\citep{he2022masked}, that have been mainly used for self-supervised pre-training of transformers~\citep{huang2023self}, can be adapted to convolutional models. Masked autoencoders divide the images into patches, mask part of the patches, and train the model to reconstruct the original images. Due to the moderate success of this method for convolutional models so far~\citep{huang2023self}, they adapted it by using sparse convolutions instead of normal convolutions for the pre-training, where they achieved comparable results to contrastive learning. In ~\citep{wolf2023self}, their method, called SparK, was applied to CT slices and shows similar performance to SwAV and MoCo for self-supervised pre-training and is particularly robust for small downstream datasets.

We compare the best-performing reduction method \hash{} with threshold six against the baseline method \all{} on the contrastive learning approach MoCo Version 2 and the masked autoencoder approach SparK. To prove that our results are generalizable to other contrastive pre-training methods, we expect to see performance gains with our slice reduction method for MoCo, since, as in SwAV, less similar images should improve the model's ability to distinguish in latent space. On the other hand, since there is no distinguishing involved in masked autoencoder pre-training approaches, similar images should not be a problem there. Thus, we expect no performance gain or slightly reduced performance for the masked autoencoder method SparK, since less similar images should not bring any advantage and the model just has less training data.
Detailed explanations of MoCo and Spark can be found in~\ref{sec:Appendix2} and~\ref{sec:Appendix3}.
Table~\ref{tab:compare} shows the downstream task results. The contrastive learning method MoCo performs better with the reduced dataset, analogous to the contrastive learning method SwAV discussed earlier. For the masked autoencoder method SparK, we do not achieve any improvements with the reduction, using all slices achieves superior results.

\begingroup
\setlength{\tabcolsep}{2.5pt}
\begin{table*}[t]
    \caption{Evaluation E: This table shows the downstream task results for pre-training with the contrastive learning (CL) approaches SwAV and MoCo Version 2 and the masked autoencoder (MAE) approach SparK. For all three approaches, we compare pre-training with all data (All) to pre-training with the reduced dataset using the hash reduction method and threshold 6 (Reduced) on the LIDC dataset. (Accuracy can be found in in Table~\ref{tab:compareApp} in~\ref{sec:Appendix5})}
    \centering
\resizebox{\textwidth}{!}{
    \begin{tabular}{l@{\hskip 4pt}l@{\hskip 7pt}cc@{\hskip 7pt}cc@{\hskip 7pt}cc}
        \toprule
         \multicolumn{2}{c}{Pre-Training} & \multicolumn{6}{c}{Downstream Results}  \\
                 \cmidrule(lr){1-2} \cmidrule(lr){3-8}             
        \multicolumn{1}{c}{Approach} & \multicolumn{1}{c}{Data} &  \multicolumn{2}{c}{\covid{}}  & \multicolumn{2}{c}{\mnist{}} & \multicolumn{2}{c}{\brain{}}  \\
        \cmidrule(lr){3-4} \cmidrule(lr){5-6} \cmidrule(lr){7-8}
        & &  AUC & F1 & AUC & F1 & AUC & F1 \\
        \midrule 
      SwAV (CL)    &  All   & 0.807 $\pm$ 0.006 & 0.744 $\pm$ 0.013 & 0.972 $\pm$ 0.003 & 0.769 $\pm$ 0.003 & 0.734 $\pm$ 0.046 & 0.609 $\pm$ 0.072 \\
              &  Reduced    & \textbf{0.823 $\pm$ 0.005}  & \textbf{0.768 $\pm$ 0.008} & \textbf{0.982 $\pm$ 0.001} & \textbf{0.802 $\pm$ 0.002} & \textbf{0.840 $\pm$ 0.016} & \textbf{0.800 $\pm$ 0.033} \\
             \rule{0pt}{3ex}%
      MoCoV2 (CL)   &  All   & 0.824 $\pm$ 0.005  & 0.780 $\pm$ 0.009   & 0.981 $\pm$ 0.001  & 0.817 $\pm$ 0.001  & 0.825 $\pm$ 0.010  & 0.770 $\pm$ 0.064\\
              &  Reduced    & \textbf{0.830 $\pm$ 0.005} & \textbf{0.781 $\pm$ 0.005} & \textbf{0.982 $\pm$ 0.003} & \textbf{0.820 $\pm$ 0.004} & \textbf{0.897 $\pm$ 0.015} & \textbf{0.791 $\pm$ 0.032} \\
                   \rule{0pt}{3ex}%
      SparK (MAE)  &  All   & \textbf{0.828 $\pm$ 0.006}  & \textbf{0.776 $\pm$ 0.009}  & \textbf{0.981 $\pm$ 0.001}  & \textbf{0.808 $\pm$ 0.003}  & \textbf{0.919 $\pm$ 0.015} & \textbf{0.812 $\pm$ 0.080}\\
              &  Reduced    & 0.810 $\pm$ 0.006 & 0.761 $\pm$ 0.020 & 0.978 $\pm$ 0.001 & 0.782 $\pm$ 0.002 & 0.882 $\pm$ 0.024 & 0.809 $\pm$ 0.051 \\

        \bottomrule
    \end{tabular}
    }
    \label{tab:compare}
\end{table*}
\endgroup

\section{Discussion}
Self-supervised pre-training of deep learning models with contrastive learning on large unannotated datasets is a common and successful approach in medical imaging to cope with small annotated datasets~\citep{huang2023self}. The most popular contrastive learning methods were initially developed for natural image processing and transferred to the medical domain~\citep{huang2023self}. Many methods can be directly applied to the medical domain without adaptation; however, not all methods show the same behavior because medical images have different structures and color schemes~\citep{raghu2019transfusion}. In this work, we investigate the composition of the pre-training datasets for contrastive learning on CT slices. We perform our investigations on two large pre-training datasets separately to ensure generalizability and evaluate the pre-trainings on three classification downstream tasks, the benchmark task for evaluating self-supervised pre-training~\citep{huang2023self}. Table~\ref{tab:NoPre} shows the results without pre-training and Table~\ref{tab:baseline}, row \pet{} ALL and row \lidc{} ALL, show the results when contrastive pre-training on all slices of the pre-training dataset is applied, the current state-of-the-art~\citep{ghesu2022contrastive,chen2021momentum}. Contrastive pre-training improves the downstream results for the \covid{} and the \brain{} tasks with both pre-training datasets. However, for the \mnist{} task, we only achieve performance gains when pre-training with the \lidc{} dataset. Contrastive pre-training on all slices of the \pet{} dataset slightly decreases the results of the \mnist{} task by 0.003 AUC score. This shows that pre-training with contrastive learning on CT scans does not improve downstream performance in all cases, which is also confirmed in Huang et al.`s.~\citep{huang2023self} study.

In contrastive learning, the model is trained to distinguish between latent space representations of positive pairs coming from two augmented views of the same original image and latent space representations of negative pairs coming from two different original images. We hypothesized that using each slice of a CT volume for contrastive pre-training might lead to a model that is unable to discriminate between positive and negative pairs since the similarity between two augmented versions of a slice might be lower than the similarity between two different slices.  In the first experiment, we reduced the pre-training datasets by using only every $n$th slice of a volume. The results listed in Table~\ref{tab:baseline} support our hypotheses, as performance improves on the downstream tasks. Using only every $n$th slice of a volume increases the variation between slices in the pre-training dataset, potentially allowing the model to better distinguish between positive and negative pairs. After the reduction, contrastive pre-training with SwAV outperforms no pre-training for all tasks and pre-training datasets, including the \mnist{} task and pre-training on the \pet{} dataset, where we had performance losses when pre-training with all slices. 

In a second experiment, we aimed to evaluate whether there are dataset reduction methods that are more suitable than the baseline reduction. Further, we aimed to find the optimal threshold for the degree of similarity between the slices that leads to the highest results in downstream tasks. As shown in Table~\ref{tab:methods}, of all evaluated methods, the \hash{} method performs best on the downstream tasks, and as shown in Table~\ref{tab:threshold}, a similarity threshold of six for the \hash{} approach seems to ensure the best degree of similarity. These experiments lead to the assumption that the \hash{} reduction approach with threshold six ensures the best compromise between a high variation and a sufficiently large number of samples in the pre-training dataset. Furthermore, with an execution time of less than 30 minutes, the \hash{} dataset reduction is computed faster than all other similarly based reduction methods evaluated. 

In a further experiment, we tested the \hash{} reduction method on two other self-supervised pre-training approaches. We expected that our results would generalize to other contrastive learning approaches due to the identical basic concept, but that reduction would not improve the results for other self-supervised pre-training methods. As shown in Table~\ref{tab:compare}, pre-training with the contrastive learning approach MoCo is improved with the \hash{} reduction method. This proves the generalizability of our results to other contrastive pre-training methods, that are trained by distinguishing between latent space representations of augmented views and original images and thus have the problem of too similar images. As also shown in Table~\ref{tab:compare} the masked autoencoder pre-training method SparK performs best with all slices. Since masked autoencoders are trained to reconstruct masked patches of images where no distinguishing is involved, reducing the dataset does not bring any advantage and the model just has less training data. The results in Table~\ref{tab:compare} support our hypothesis that selective CT data reduction is beneficial for contrastive pre-training due to the distinguishing challenge, but pre-training methods that do not rely on distinguishing in latent space do not benefit from dataset reduction.

A major benefit of pre-training dataset reduction for contrastive learning is that we significantly reduce the pre-training time. With less time and thus less energy cost, better pre-training results on CT image classification downstream tasks can be achieved, as summarized in Figure~\ref{fig:results}.

\section{Limitations}
As our work shows the great potential of CT dataset reduction for contrastive pre-training, it would be interesting to further investigate these findings in future research.  A limitation of our work is that we only chose well-established, computationally fast dataset reduction methods that are based on similarity calculations. For future work, it would be interesting to see if there are other dataset reduction techniques that could lead to even better results. For example, core-set selection such as SVP (Selection via Proxy)~\citep{Coleman2020Selection} or CRAIG (Coresets for Accelerating Incremental Gradient descent)~\citep{mirzasoleiman2020coresets} might be a promising idea. However, these methods require significantly more time and computational effort for the dataset reduction.   

Another limitation of our work is that we did not investigate different augmentation strategies for the contrastive pre-training. The ability to distinguish between latent space representations coming from two augmented views of the same original image and latent space representations coming from two different original images also depends on the type and amount of augmentations used. We took exactly the augmentations of the original contrastive learning publications SwAV~\citep{caron2020unsupervised} and MoCo~\citep{he2020momentum}, since they have done an intensive evaluation on which augmentation techniques are best suited for their pre-training method. However, for future work, it would be interesting to investigate different types and amounts of augmentation to see if less data reduction is needed and if even better results can be achieved. Furthermore, it would be interesting to better understand why exactly the \hash{} method leads to the best results and if this is dependent on the used augmentation methods.    

A further limitation is that our experiments were performed with only one deep learning model. Analogous to the original publications of the two self-supervised pre-training methods SwAV~\citep{caron2020unsupervised} and MoCo~\citep{he2020momentum}, we chose the ResNet50~\citep{he2016deep} as our model, due to its widespread use as a baseline for comparisons in vision studies~\citep{liu2022convnet} that were later successfully transferred to other models and its popularity in medical image analysis~\citep{kora2022transfer}. For further research, it would be interesting to apply our findings to other deep learning models as well. Furthermore, applying our results to other modalities of volumetric images consisting of consecutive slices like MRI or PET, would be possible for future research projects. 

In addition to the very important task of CT image classification with a lot of ongoing research~\citep{kim2022transfer,huang2023self} CT image segmentation is another popular task in medical imaging. We have tested our self-supervised pre-trained classification model on several segmentation tasks by adding a U-Net~\citep{ronneberger2015u} decoder to the pre-trained ResNet50. However, we did not achieve any performance gains with our pre-training, neither on \all{} data nor on \hash{}-6 reduced data. Thus, a clear limitation of our work is that our results cannot be directly applied to downstream segmentation tasks. As shown in~\citep{wu2024voco}, pure contrastive pre-training with methods from the RGB imaging domain only on the encoder does not lead to significant performance gains for segmentation downstream tasks. Instead, other specific pre-training methods for segmentation can lead to improved performance. For future work, it would be interesting to further investigate our findings on such segmentation-optimized contrastive learning methods.   

Another idea would be to combine our findings with the work of Joshua, et al.~\citep{robinson2021contrastive}. With our proposed \hash{}-6 approach, slices that are too similar to be distinguished by the model when using contrastive pre-training can be identified and removed from the dataset before the training. Joshua, et al. developed a method to improve the model`s pre-training, by targeting the samples that turn out to be difficult for the model to distinguish during the training. This is done by analyzing in the latent space which images the model places close together, but should actually be far apart, as they are negative samples that should be pushed apart in the latent space. These samples, called hard negative samples, are pushed apart by a special loss function. So one possibility would be to first apply our approach to filter out images that are too similar to be meaningfully distinguished. Then, hard negative mining could be applied to further improve the pre-training by targeting cases that turn out to be still difficult for the model during the training. Hard negative mining by Joshua, et al. has originally been developed on RGB images. We see strong potential in adapting this method to CT slices and exploring the combination of the two approaches.

By performing pre-training and fine-tuning with a 2D model on the slices of CT scans instead of using a 3D model on volumes, we ensure low computational costs for inference on downstream tasks so our findings can be applied globally without requiring powerful GPUs. However, training 3D models on volumes and training 2D models on slices are both widely used approaches for deep learning on CT scans, with several recent publications demonstrating excellent results for clinically relevant CT imaging tasks on both 3D~\citep{lisson2022deep,andrearczyk2021overview}, and 2D~\citep{wang2021deep,jiang2022dynamic,xing2022cs,baghdadi2022automated} models. Both approaches have their advantages. After evaluating the pre-training dataset composition for contrastive pre-training of 2D models on CT slices, for future work, an evaluation of the properties of pre-training datasets with entire volumes would be interesting.

\section{Conclusion}
In our work, we investigate how to exploit the characteristics of CT datasets to improve contrastive pre-training. We hypothesized that using all slices in each CT volume of a pre-training dataset may lead to performance degradation due to the low variation in the data. The experiments, with over 2000 pre-training hours, support our hypothesis. In conclusion, we propose to reduce pre-training datasets using the \hash{} method and a threshold of six. This approach leads to considerable performance gains in classification downstream tasks in all our experiments and outperforms the other evaluated dataset reduction methods. The time to reduce the datasets using the \hash{} approach is negligibly short, with execution times of less than half an hour, while the pre-training duration is substantially reduced. Research on CT data with contrastive learning in the future can incorporate our findings to improve their performance on classification tasks and speed up learning by reducing their pre-training dataset with our proposed method.

\section*{CRediT authorship contribution statement}
Conceptualization: D.W.; Methodology: D.W., T.P.; Software: D.W., T.P; Validation: D.W., M.G, T.R.; Formal analysis D.W.; Investigation: D.W; Resources: M.G, T.R., M.B.; Data Curation: C.S.L., C.G.L; Writing - original draft: D.W.; Writing - review and editing: D.W., T.P., M.G, T.R., C.S.L., C.G.L, M.B.; Visualization: D.W.; Supervision: M.G, T.R., M.B.; Project administration:  M.G, T.R., M.B.; Funding acquisition: M.B.

\section*{Declaration of Competing Interest}
The authors declare no competing interests.

\section*{Data availability}
Pre-Training: The LIDC-IDRI~\citep{armato2011lung,armato2015lung} dataset is available for public use under the license CC BY 3.0. The FDG-PET-CT~\citep{gatidis2022data,gatidis2022whole} dataset is available for public use listed under request at TCIA~\citep{clark2013cancer}.

Downstream: The COVID-19 CT Classification Grand Challenge~\cite{yang2020covid} dataset is available at \url{https://github.com/UCSD-AI4H/COVID-CT}; The OrganSMNIST dataset from MedMNIST~\cite{medmnistv2} is available for public use under the license CC BY 4.0; The internal \brain{} dataset cannot be made publicly available due to strict data security restrictions. 

Code will be released in case of acceptance.

\section*{Acknowledgments}
The authors acknowledge the National Cancer Institute and the Foundation for the National Institutes of Health, and their critical role in the creation of the free publicly available LIDC/IDRI Database used in this study.

\section*{Funding}
This work is funded by ``NUM 2.0''  (FKZ: 01KX2121) as part of the Racoon Project.

\section*{Ethics declarations}
For the internal \brain{} task, ethical approval was granted by the Ethics Committee of Ulm University under ID 302/17. The procedure was in accordance with the ethical standards of the World Medical Association (Declaration of Helsinki).

\bibliographystyle{elsarticle-harv} 
\bibliography{cas-refs}

\begin{thebibliography}{68}
\expandafter\ifx\csname natexlab\endcsname\relax\def\natexlab#1{#1}\fi
\providecommand{\url}[1]{\texttt{#1}}
\providecommand{\href}[2]{#2}
\providecommand{\path}[1]{#1}
\providecommand{\DOIprefix}{doi:}
\providecommand{\ArXivprefix}{arXiv:}
\providecommand{\URLprefix}{URL: }
\providecommand{\Pubmedprefix}{pmid:}
\providecommand{\doi}[1]{\href{http://dx.doi.org/#1}{\path{#1}}}
\providecommand{\Pubmed}[1]{\href{pmid:#1}{\path{#1}}}
\providecommand{\bibinfo}[2]{#2}
\ifx\xfnm\relax \def\xfnm[#1]{\unskip,\space#1}\fi
\bibitem[{Andrearczyk et~al.(2021)Andrearczyk, Oreiller, Jreige, Vallieres, Castelli, Elhalawani, Boughdad, Prior and Depeursinge}]{andrearczyk2021overview}
\bibinfo{author}{Andrearczyk, V.}, \bibinfo{author}{Oreiller, V.}, \bibinfo{author}{Jreige, M.}, \bibinfo{author}{Vallieres, M.}, \bibinfo{author}{Castelli, J.}, \bibinfo{author}{Elhalawani, H.}, \bibinfo{author}{Boughdad, S.}, \bibinfo{author}{Prior, J.O.}, \bibinfo{author}{Depeursinge, A.}, \bibinfo{year}{2021}.
\newblock \bibinfo{title}{Overview of the hecktor challenge at miccai 2020: automatic head and neck tumor segmentation in pet/ct}, in: \bibinfo{booktitle}{Head and Neck Tumor Segmentation: First Challenge, HECKTOR 2020, Held in Conjunction with MICCAI 2020, Lima, Peru, October 4, 2020, Proceedings 1}, \bibinfo{organization}{Springer}. pp. \bibinfo{pages}{1--21}.
\bibitem[{Armato~III et~al.(2011)}]{armato2011lung}
\bibinfo{author}{Armato~III, S.G.}, et~al., \bibinfo{year}{2011}.
\newblock \bibinfo{title}{The lung image database consortium (lidc) and image database resource initiative (idri): a completed reference database of lung nodules on ct scans}.
\newblock \bibinfo{journal}{Medical physics} \bibinfo{volume}{38}, \bibinfo{pages}{915--931}.
\bibitem[{Armato~III et~al.(2015)}]{armato2015lung}
\bibinfo{author}{Armato~III, S.G.}, et~al., \bibinfo{year}{2015}.
\newblock \bibinfo{title}{Data from lidc-idri (data set)}.
\newblock \bibinfo{journal}{The Cancer Imaging Archive.} .
\bibitem[{Avesta et~al.(2023)Avesta, Hossain, Lin, Aboian, Krumholz and Aneja}]{avesta2023comparing}
\bibinfo{author}{Avesta, A.}, \bibinfo{author}{Hossain, S.}, \bibinfo{author}{Lin, M.}, \bibinfo{author}{Aboian, M.}, \bibinfo{author}{Krumholz, H.M.}, \bibinfo{author}{Aneja, S.}, \bibinfo{year}{2023}.
\newblock \bibinfo{title}{Comparing 3d, 2.5 d, and 2d approaches to brain image auto-segmentation}.
\newblock \bibinfo{journal}{Bioengineering} \bibinfo{volume}{10}, \bibinfo{pages}{181}.
\bibitem[{Azizi et~al.(2021)Azizi, Mustafa, Ryan, Beaver, Freyberg, Deaton, Loh, Karthikesalingam, Kornblith, Chen et~al.}]{azizi2021big}
\bibinfo{author}{Azizi, S.}, \bibinfo{author}{Mustafa, B.}, \bibinfo{author}{Ryan, F.}, \bibinfo{author}{Beaver, Z.}, \bibinfo{author}{Freyberg, J.}, \bibinfo{author}{Deaton, J.}, \bibinfo{author}{Loh, A.}, \bibinfo{author}{Karthikesalingam, A.}, \bibinfo{author}{Kornblith, S.}, \bibinfo{author}{Chen, T.}, et~al., \bibinfo{year}{2021}.
\newblock \bibinfo{title}{Big self-supervised models advance medical image classification}, in: \bibinfo{booktitle}{Proceedings of the IEEE/CVF international conference on computer vision}, pp. \bibinfo{pages}{3478--3488}.
\bibitem[{Baghdadi et~al.(2022)Baghdadi, Malki, Abdelaliem, Balaha, Badawy and Elhosseini}]{baghdadi2022automated}
\bibinfo{author}{Baghdadi, N.A.}, \bibinfo{author}{Malki, A.}, \bibinfo{author}{Abdelaliem, S.F.}, \bibinfo{author}{Balaha, H.M.}, \bibinfo{author}{Badawy, M.}, \bibinfo{author}{Elhosseini, M.}, \bibinfo{year}{2022}.
\newblock \bibinfo{title}{An automated diagnosis and classification of covid-19 from chest ct images using a transfer learning-based convolutional neural network}.
\newblock \bibinfo{journal}{Computers in biology and medicine} \bibinfo{volume}{144}, \bibinfo{pages}{105383}.
\bibitem[{Balestriero et~al.(2023)Balestriero, Ibrahim, Sobal, Morcos, Shekhar, Goldstein, Bordes, Bardes, Mialon, Tian et~al.}]{balestriero2023cookbook}
\bibinfo{author}{Balestriero, R.}, \bibinfo{author}{Ibrahim, M.}, \bibinfo{author}{Sobal, V.}, \bibinfo{author}{Morcos, A.}, \bibinfo{author}{Shekhar, S.}, \bibinfo{author}{Goldstein, T.}, \bibinfo{author}{Bordes, F.}, \bibinfo{author}{Bardes, A.}, \bibinfo{author}{Mialon, G.}, \bibinfo{author}{Tian, Y.}, et~al., \bibinfo{year}{2023}.
\newblock \bibinfo{title}{A cookbook of self-supervised learning}.
\newblock \bibinfo{journal}{arXiv preprint arXiv:2304.12210} .
\bibitem[{Bhattacharjee et~al.(2021)Bhattacharjee, Douglas, Drukker, Hu, Fuhrman, Sheth and Giger}]{bhattacharjee2021comparison}
\bibinfo{author}{Bhattacharjee, R.}, \bibinfo{author}{Douglas, L.}, \bibinfo{author}{Drukker, K.}, \bibinfo{author}{Hu, Q.}, \bibinfo{author}{Fuhrman, J.}, \bibinfo{author}{Sheth, D.}, \bibinfo{author}{Giger, M.}, \bibinfo{year}{2021}.
\newblock \bibinfo{title}{Comparison of 2d and 3d u-net breast lesion segmentations on dce-mri}, in: \bibinfo{booktitle}{Medical Imaging 2021: Computer-Aided Diagnosis}, \bibinfo{organization}{SPIE}. pp. \bibinfo{pages}{81--87}.
\bibitem[{Caron et~al.(2020)Caron, Misra, Mairal, Goyal, Bojanowski and Joulin}]{caron2020unsupervised}
\bibinfo{author}{Caron, M.}, \bibinfo{author}{Misra, I.}, \bibinfo{author}{Mairal, J.}, \bibinfo{author}{Goyal, P.}, \bibinfo{author}{Bojanowski, P.}, \bibinfo{author}{Joulin, A.}, \bibinfo{year}{2020}.
\newblock \bibinfo{title}{Unsupervised learning of visual features by contrasting cluster assignments}.
\newblock \bibinfo{journal}{Advances in neural information processing systems} \bibinfo{volume}{33}, \bibinfo{pages}{9912--9924}.
\bibitem[{Chen et~al.(2020a)Chen, Kornblith, Norouzi and Hinton}]{chen2020simple}
\bibinfo{author}{Chen, T.}, \bibinfo{author}{Kornblith, S.}, \bibinfo{author}{Norouzi, M.}, \bibinfo{author}{Hinton, G.}, \bibinfo{year}{2020}a.
\newblock \bibinfo{title}{A simple framework for contrastive learning of visual representations}, in: \bibinfo{booktitle}{International conference on machine learning}, \bibinfo{organization}{PMLR}. pp. \bibinfo{pages}{1597--1607}.
\bibitem[{Chen et~al.(2020b)Chen, Fan, Girshick and He}]{chen2020improved}
\bibinfo{author}{Chen, X.}, \bibinfo{author}{Fan, H.}, \bibinfo{author}{Girshick, R.}, \bibinfo{author}{He, K.}, \bibinfo{year}{2020}b.
\newblock \bibinfo{title}{Improved baselines with momentum contrastive learning}.
\newblock \bibinfo{journal}{arXiv preprint arXiv:2003.04297} .
\bibitem[{Chen et~al.(2021)Chen, Yao, Zhou, Dong and Zhang}]{chen2021momentum}
\bibinfo{author}{Chen, X.}, \bibinfo{author}{Yao, L.}, \bibinfo{author}{Zhou, T.}, \bibinfo{author}{Dong, J.}, \bibinfo{author}{Zhang, Y.}, \bibinfo{year}{2021}.
\newblock \bibinfo{title}{Momentum contrastive learning for few-shot covid-19 diagnosis from chest ct images}.
\newblock \bibinfo{journal}{Pattern recognition} \bibinfo{volume}{113}, \bibinfo{pages}{107826}.
\bibitem[{Clark(2015)}]{clark2015pillow}
\bibinfo{author}{Clark, A.}, \bibinfo{year}{2015}.
\newblock \bibinfo{title}{Pillow (pil fork) documentation [acsession date: 2023-07-01]}.
\newblock \URLprefix \url{https://buildmedia.readthedocs.org/media/pdf/pillow/latest/pillow.pdf}.
\bibitem[{Clark et~al.(2013)}]{clark2013cancer}
\bibinfo{author}{Clark, K.}, et~al., \bibinfo{year}{2013}.
\newblock \bibinfo{title}{The cancer imaging archive (tcia): maintaining and operating a public information repository}.
\newblock \bibinfo{journal}{Journal of digital imaging} \bibinfo{volume}{26}, \bibinfo{pages}{1045--1057}.
\bibitem[{Coleman et~al.(2020)Coleman, Yeh, Mussmann, Mirzasoleiman, Bailis, Liang, Leskovec and Zaharia}]{Coleman2020Selection}
\bibinfo{author}{Coleman, C.}, \bibinfo{author}{Yeh, C.}, \bibinfo{author}{Mussmann, S.}, \bibinfo{author}{Mirzasoleiman, B.}, \bibinfo{author}{Bailis, P.}, \bibinfo{author}{Liang, P.}, \bibinfo{author}{Leskovec, J.}, \bibinfo{author}{Zaharia, M.}, \bibinfo{year}{2020}.
\newblock \bibinfo{title}{Selection via proxy: Efficient data selection for deep learning}, in: \bibinfo{booktitle}{International Conference on Learning Representations}.
\bibitem[{Conrad and Narayan(2021)}]{conrad2021cem500k}
\bibinfo{author}{Conrad, R.}, \bibinfo{author}{Narayan, K.}, \bibinfo{year}{2021}.
\newblock \bibinfo{title}{Cem500k, a large-scale heterogeneous unlabeled cellular electron microscopy image dataset for deep learning}.
\newblock \bibinfo{journal}{Elife} \bibinfo{volume}{10}, \bibinfo{pages}{e65894}.
\bibitem[{Consortium(2022)}]{monai_consortium_2022_7086266}
\bibinfo{author}{Consortium, M.}, \bibinfo{year}{2022}.
\newblock \bibinfo{title}{Monai: Medical open network for ai: 1.0.0 release [acsession date: 2023-07-01]}.
\newblock \DOIprefix\doi{10.5281/zenodo.7086266}.
\bibitem[{Cuturi(2013)}]{cuturi2013sinkhorn}
\bibinfo{author}{Cuturi, M.}, \bibinfo{year}{2013}.
\newblock \bibinfo{title}{Sinkhorn distances: Lightspeed computation of optimal transport}.
\newblock \bibinfo{journal}{Advances in neural information processing systems} \bibinfo{volume}{26}.
\bibitem[{Deng et~al.(2009)Deng, Dong, Socher, Li, Li and Fei-Fei}]{deng2009imagenet}
\bibinfo{author}{Deng, J.}, \bibinfo{author}{Dong, W.}, \bibinfo{author}{Socher, R.}, \bibinfo{author}{Li, L.J.}, \bibinfo{author}{Li, K.}, \bibinfo{author}{Fei-Fei, L.}, \bibinfo{year}{2009}.
\newblock \bibinfo{title}{Imagenet: A large-scale hierarchical image database}, in: \bibinfo{booktitle}{2009 IEEE conference on computer vision and pattern recognition}, \bibinfo{organization}{Ieee}. pp. \bibinfo{pages}{248--255}.
\bibitem[{Dufumier et~al.(2021)Dufumier, Gori, Victor, Grigis, Wessa, Brambilla, Favre, Polosan, McDonald, Piguet et~al.}]{dufumier2021contrastive}
\bibinfo{author}{Dufumier, B.}, \bibinfo{author}{Gori, P.}, \bibinfo{author}{Victor, J.}, \bibinfo{author}{Grigis, A.}, \bibinfo{author}{Wessa, M.}, \bibinfo{author}{Brambilla, P.}, \bibinfo{author}{Favre, P.}, \bibinfo{author}{Polosan, M.}, \bibinfo{author}{McDonald, C.}, \bibinfo{author}{Piguet, C.M.}, et~al., \bibinfo{year}{2021}.
\newblock \bibinfo{title}{Contrastive learning with continuous proxy meta-data for 3d mri classification}, in: \bibinfo{booktitle}{Medical Image Computing and Computer Assisted Intervention--MICCAI 2021: 24th International Conference, Strasbourg, France, September 27--October 1, 2021, Proceedings, Part II 24}, \bibinfo{organization}{Springer}. pp. \bibinfo{pages}{58--68}.
\bibitem[{Ewen and Khan(2021)}]{ewen2021targeted}
\bibinfo{author}{Ewen, N.}, \bibinfo{author}{Khan, N.}, \bibinfo{year}{2021}.
\newblock \bibinfo{title}{Targeted self supervision for classification on a small covid-19 ct scan dataset}, in: \bibinfo{booktitle}{2021 IEEE 18th International Symposium on Biomedical Imaging (ISBI)}, \bibinfo{organization}{IEEE}. pp. \bibinfo{pages}{1481--1485}.
\bibitem[{Falcon et~al.(2020)Falcon, Borovec, W{\"a}lchli, Eggert, Schock, Jordan, Skafte, Bereznyuk, Harris, Murrell et~al.}]{william_falcon_2020_3828935}
\bibinfo{author}{Falcon, W.}, \bibinfo{author}{Borovec, J.}, \bibinfo{author}{W{\"a}lchli, A.}, \bibinfo{author}{Eggert, N.}, \bibinfo{author}{Schock, J.}, \bibinfo{author}{Jordan, J.}, \bibinfo{author}{Skafte, N.}, \bibinfo{author}{Bereznyuk, V.}, \bibinfo{author}{Harris, E.}, \bibinfo{author}{Murrell, T.}, et~al., \bibinfo{year}{2020}.
\newblock \bibinfo{title}{Pytorchlightning/pytorch-lightning: 0.7.6 release [acsession date: 2023-07-01]}.
\newblock \DOIprefix\doi{10.5281/zenodo.3828935}.
\bibitem[{Gatidis and K{\"u}stner(2022)}]{gatidis2022data}
\bibinfo{author}{Gatidis, S.}, \bibinfo{author}{K{\"u}stner, T.}, \bibinfo{year}{2022}.
\newblock \bibinfo{title}{A whole-body fdg-pet/ct dataset with manually annotated tumor lesions (fdg-pet-ct-lesions) [dataset]}.
\newblock \bibinfo{journal}{The Cancer Imaging Archive} .
\bibitem[{Gatidis et~al.(2022a)Gatidis, Küstner, Ingrisch, Fabritius and Cyran}]{sergios_gatidis_2022_6362493}
\bibinfo{author}{Gatidis, S.}, \bibinfo{author}{Küstner, T.}, \bibinfo{author}{Ingrisch, M.}, \bibinfo{author}{Fabritius, M.}, \bibinfo{author}{Cyran, C.}, \bibinfo{year}{2022}a.
\newblock \bibinfo{title}{{Automated Lesion Segmentation in Whole-Body FDG- PET/CT} [acsession date: 2023-07-01]}.
\newblock \DOIprefix\doi{10.5281/zenodo.6362493}.
\bibitem[{Gatidis et~al.(2022b)}]{gatidis2022whole}
\bibinfo{author}{Gatidis, S.}, et~al., \bibinfo{year}{2022}b.
\newblock \bibinfo{title}{A whole-body fdg-pet/ct dataset with manually annotated tumor lesions}.
\newblock \bibinfo{journal}{Scientific Data} \bibinfo{volume}{9}, \bibinfo{pages}{601}.
\bibitem[{Ghesu et~al.(2022)Ghesu, Georgescu, Mansoor, Yoo, Neumann, Patel, Vishwanath, Balter, Cao, Grbic et~al.}]{ghesu2022contrastive}
\bibinfo{author}{Ghesu, F.C.}, \bibinfo{author}{Georgescu, B.}, \bibinfo{author}{Mansoor, A.}, \bibinfo{author}{Yoo, Y.}, \bibinfo{author}{Neumann, D.}, \bibinfo{author}{Patel, P.}, \bibinfo{author}{Vishwanath, R.S.}, \bibinfo{author}{Balter, J.M.}, \bibinfo{author}{Cao, Y.}, \bibinfo{author}{Grbic, S.}, et~al., \bibinfo{year}{2022}.
\newblock \bibinfo{title}{Contrastive self-supervised learning from 100 million medical images with optional supervision}.
\newblock \bibinfo{journal}{Journal of Medical Imaging} \bibinfo{volume}{9}, \bibinfo{pages}{064503--064503}.
\bibitem[{Gildenblat and contributors(2021)}]{jacobgilpytorchcam}
\bibinfo{author}{Gildenblat, J.}, \bibinfo{author}{contributors}, \bibinfo{year}{2021}.
\newblock \bibinfo{title}{Pytorch library for {CAM} methods [acsession date: 2023-07-01]}.
\newblock \URLprefix \url{https://github.com/jacobgil/pytorch-grad-cam}.
\bibitem[{Grill et~al.(2020)Grill, Strub, Altch{\'e}, Tallec, Richemond, Buchatskaya, Doersch, Avila~Pires, Guo, Gheshlaghi~Azar et~al.}]{grill2020bootstrap}
\bibinfo{author}{Grill, J.B.}, \bibinfo{author}{Strub, F.}, \bibinfo{author}{Altch{\'e}, F.}, \bibinfo{author}{Tallec, C.}, \bibinfo{author}{Richemond, P.}, \bibinfo{author}{Buchatskaya, E.}, \bibinfo{author}{Doersch, C.}, \bibinfo{author}{Avila~Pires, B.}, \bibinfo{author}{Guo, Z.}, \bibinfo{author}{Gheshlaghi~Azar, M.}, et~al., \bibinfo{year}{2020}.
\newblock \bibinfo{title}{Bootstrap your own latent-a new approach to self-supervised learning}.
\newblock \bibinfo{journal}{Advances in neural information processing systems} \bibinfo{volume}{33}, \bibinfo{pages}{21271--21284}.
\bibitem[{He et~al.(2022)He, Chen, Xie, Li, Doll{\'a}r and Girshick}]{he2022masked}
\bibinfo{author}{He, K.}, \bibinfo{author}{Chen, X.}, \bibinfo{author}{Xie, S.}, \bibinfo{author}{Li, Y.}, \bibinfo{author}{Doll{\'a}r, P.}, \bibinfo{author}{Girshick, R.}, \bibinfo{year}{2022}.
\newblock \bibinfo{title}{Masked autoencoders are scalable vision learners}, in: \bibinfo{booktitle}{Proceedings of the IEEE/CVF conference on computer vision and pattern recognition}, pp. \bibinfo{pages}{16000--16009}.
\bibitem[{He et~al.(2020)He, Fan, Wu, Xie and Girshick}]{he2020momentum}
\bibinfo{author}{He, K.}, \bibinfo{author}{Fan, H.}, \bibinfo{author}{Wu, Y.}, \bibinfo{author}{Xie, S.}, \bibinfo{author}{Girshick, R.}, \bibinfo{year}{2020}.
\newblock \bibinfo{title}{Momentum contrast for unsupervised visual representation learning}, in: \bibinfo{booktitle}{Proceedings of the IEEE/CVF conference on computer vision and pattern recognition}, pp. \bibinfo{pages}{9729--9738}.
\bibitem[{He et~al.(2016)He, Zhang, Ren and Sun}]{he2016deep}
\bibinfo{author}{He, K.}, \bibinfo{author}{Zhang, X.}, \bibinfo{author}{Ren, S.}, \bibinfo{author}{Sun, J.}, \bibinfo{year}{2016}.
\newblock \bibinfo{title}{Deep residual learning for image recognition}, in: \bibinfo{booktitle}{Proceedings of the IEEE conference on computer vision and pattern recognition}, pp. \bibinfo{pages}{770--778}.
\bibitem[{Huang et~al.(2023)Huang, Pareek, Jensen, Lungren, Yeung and Chaudhari}]{huang2023self}
\bibinfo{author}{Huang, S.C.}, \bibinfo{author}{Pareek, A.}, \bibinfo{author}{Jensen, M.}, \bibinfo{author}{Lungren, M.P.}, \bibinfo{author}{Yeung, S.}, \bibinfo{author}{Chaudhari, A.S.}, \bibinfo{year}{2023}.
\newblock \bibinfo{title}{Self-supervised learning for medical image classification: a systematic review and implementation guidelines}.
\newblock \bibinfo{journal}{NPJ Digital Medicine} \bibinfo{volume}{6}, \bibinfo{pages}{74}.
\bibitem[{Jiang et~al.(2022)Jiang, Yang, Li, Liu, Heng and Dou}]{jiang2022dynamic}
\bibinfo{author}{Jiang, M.}, \bibinfo{author}{Yang, H.}, \bibinfo{author}{Li, X.}, \bibinfo{author}{Liu, Q.}, \bibinfo{author}{Heng, P.A.}, \bibinfo{author}{Dou, Q.}, \bibinfo{year}{2022}.
\newblock \bibinfo{title}{Dynamic bank learning for semi-supervised federated image diagnosis with class imbalance}, in: \bibinfo{booktitle}{International Conference on Medical Image Computing and Computer-Assisted Intervention}, \bibinfo{organization}{Springer}. pp. \bibinfo{pages}{196--206}.
\bibitem[{Jing et~al.(2022)Jing, Vincent, LeCun and Tian}]{jing2022understanding}
\bibinfo{author}{Jing, L.}, \bibinfo{author}{Vincent, P.}, \bibinfo{author}{LeCun, Y.}, \bibinfo{author}{Tian, Y.}, \bibinfo{year}{2022}.
\newblock \bibinfo{title}{Understanding dimensional collapse in contrastive self-supervised learning}, in: \bibinfo{booktitle}{International Conference on Learning Representations}.
\bibitem[{Kern et~al.(2021)Kern, Klauck, Ropinski and Mastmeyer}]{kern20212d}
\bibinfo{author}{Kern, D.}, \bibinfo{author}{Klauck, U.}, \bibinfo{author}{Ropinski, T.}, \bibinfo{author}{Mastmeyer, A.}, \bibinfo{year}{2021}.
\newblock \bibinfo{title}{2d vs. 3d u-net abdominal organ segmentation in ct data using organ bounds}, in: \bibinfo{booktitle}{Medical Imaging 2021: Imaging Informatics for Healthcare, Research, and Applications}, \bibinfo{organization}{SPIE}. pp. \bibinfo{pages}{192--200}.
\bibitem[{Kim and Han(2023)}]{kim2023stability}
\bibinfo{author}{Kim, D.}, \bibinfo{author}{Han, B.}, \bibinfo{year}{2023}.
\newblock \bibinfo{title}{On the stability-plasticity dilemma of class-incremental learning}, in: \bibinfo{booktitle}{Proceedings of the IEEE/CVF Conference on Computer Vision and Pattern Recognition}, pp. \bibinfo{pages}{20196--20204}.
\bibitem[{Kim et~al.(2022)Kim, Cosa-Linan, Santhanam, Jannesari, Maros and Ganslandt}]{kim2022transfer}
\bibinfo{author}{Kim, H.E.}, \bibinfo{author}{Cosa-Linan, A.}, \bibinfo{author}{Santhanam, N.}, \bibinfo{author}{Jannesari, M.}, \bibinfo{author}{Maros, M.E.}, \bibinfo{author}{Ganslandt, T.}, \bibinfo{year}{2022}.
\newblock \bibinfo{title}{Transfer learning for medical image classification: a literature review}.
\newblock \bibinfo{journal}{BMC medical imaging} \bibinfo{volume}{22}, \bibinfo{pages}{69}.
\bibitem[{Kiryati and Landau(2021)}]{kiryati2021dataset}
\bibinfo{author}{Kiryati, N.}, \bibinfo{author}{Landau, Y.}, \bibinfo{year}{2021}.
\newblock \bibinfo{title}{Dataset growth in medical image analysis research}.
\newblock \bibinfo{journal}{Journal of imaging} \bibinfo{volume}{7}, \bibinfo{pages}{155}.
\bibitem[{Kora et~al.(2022)Kora, Ooi, Faust, Raghavendra, Gudigar, Chan, Meenakshi, Swaraja, Plawiak and Acharya}]{kora2022transfer}
\bibinfo{author}{Kora, P.}, \bibinfo{author}{Ooi, C.P.}, \bibinfo{author}{Faust, O.}, \bibinfo{author}{Raghavendra, U.}, \bibinfo{author}{Gudigar, A.}, \bibinfo{author}{Chan, W.Y.}, \bibinfo{author}{Meenakshi, K.}, \bibinfo{author}{Swaraja, K.}, \bibinfo{author}{Plawiak, P.}, \bibinfo{author}{Acharya, U.R.}, \bibinfo{year}{2022}.
\newblock \bibinfo{title}{Transfer learning techniques for medical image analysis: A review}.
\newblock \bibinfo{journal}{Biocybernetics and Biomedical Engineering} \bibinfo{volume}{42}, \bibinfo{pages}{79--107}.
\bibitem[{Kornblith et~al.(2019)Kornblith, Norouzi, Lee and Hinton}]{kornblith2019similarity}
\bibinfo{author}{Kornblith, S.}, \bibinfo{author}{Norouzi, M.}, \bibinfo{author}{Lee, H.}, \bibinfo{author}{Hinton, G.}, \bibinfo{year}{2019}.
\newblock \bibinfo{title}{Similarity of neural network representations revisited}, in: \bibinfo{booktitle}{International conference on machine learning}, \bibinfo{organization}{PMLR}. pp. \bibinfo{pages}{3519--3529}.
\bibitem[{Lisson et~al.(2022)Lisson, Lisson, Mezger, Wolf, Schmidt, Thaiss, Tausch, Beer, Stilgenbauer, Beer et~al.}]{lisson2022deep}
\bibinfo{author}{Lisson, C.S.}, \bibinfo{author}{Lisson, C.G.}, \bibinfo{author}{Mezger, M.F.}, \bibinfo{author}{Wolf, D.}, \bibinfo{author}{Schmidt, S.A.}, \bibinfo{author}{Thaiss, W.M.}, \bibinfo{author}{Tausch, E.}, \bibinfo{author}{Beer, A.J.}, \bibinfo{author}{Stilgenbauer, S.}, \bibinfo{author}{Beer, M.}, et~al., \bibinfo{year}{2022}.
\newblock \bibinfo{title}{Deep neural networks and machine learning radiomics modelling for prediction of relapse in mantle cell lymphoma}.
\newblock \bibinfo{journal}{Cancers} \bibinfo{volume}{14}, \bibinfo{pages}{2008}.
\bibitem[{Liu et~al.(2022)Liu, Mao, Wu, Feichtenhofer, Darrell and Xie}]{liu2022convnet}
\bibinfo{author}{Liu, Z.}, \bibinfo{author}{Mao, H.}, \bibinfo{author}{Wu, C.Y.}, \bibinfo{author}{Feichtenhofer, C.}, \bibinfo{author}{Darrell, T.}, \bibinfo{author}{Xie, S.}, \bibinfo{year}{2022}.
\newblock \bibinfo{title}{A convnet for the 2020s}, in: \bibinfo{booktitle}{Proceedings of the IEEE/CVF conference on computer vision and pattern recognition}, pp. \bibinfo{pages}{11976--11986}.
\bibitem[{Van~der Maaten and Hinton(2008)}]{van2008visualizing}
\bibinfo{author}{Van~der Maaten, L.}, \bibinfo{author}{Hinton, G.}, \bibinfo{year}{2008}.
\newblock \bibinfo{title}{Visualizing data using t-sne.}
\newblock \bibinfo{journal}{Journal of machine learning research} \bibinfo{volume}{9}.
\bibitem[{Maier-Hein et~al.(2018)Maier-Hein, Eisenmann, Reinke, Onogur, Stankovic, Scholz, Arbel, Bogunovic, Bradley, Carass et~al.}]{maier2018rankings}
\bibinfo{author}{Maier-Hein, L.}, \bibinfo{author}{Eisenmann, M.}, \bibinfo{author}{Reinke, A.}, \bibinfo{author}{Onogur, S.}, \bibinfo{author}{Stankovic, M.}, \bibinfo{author}{Scholz, P.}, \bibinfo{author}{Arbel, T.}, \bibinfo{author}{Bogunovic, H.}, \bibinfo{author}{Bradley, A.P.}, \bibinfo{author}{Carass, A.}, et~al., \bibinfo{year}{2018}.
\newblock \bibinfo{title}{Why rankings of biomedical image analysis competitions should be interpreted with care}.
\newblock \bibinfo{journal}{Nature communications} \bibinfo{volume}{9}, \bibinfo{pages}{5217}.
\bibitem[{Mirzasoleiman et~al.(2020)Mirzasoleiman, Bilmes and Leskovec}]{mirzasoleiman2020coresets}
\bibinfo{author}{Mirzasoleiman, B.}, \bibinfo{author}{Bilmes, J.}, \bibinfo{author}{Leskovec, J.}, \bibinfo{year}{2020}.
\newblock \bibinfo{title}{Coresets for data-efficient training of machine learning models}, in: \bibinfo{booktitle}{International Conference on Machine Learning}, \bibinfo{organization}{PMLR}. pp. \bibinfo{pages}{6950--6960}.
\bibitem[{Nemoto et~al.(2020)Nemoto, Futakami, Yagi, Kumabe, Takeda, Kunieda and Shigematsu}]{nemoto2020efficacy}
\bibinfo{author}{Nemoto, T.}, \bibinfo{author}{Futakami, N.}, \bibinfo{author}{Yagi, M.}, \bibinfo{author}{Kumabe, A.}, \bibinfo{author}{Takeda, A.}, \bibinfo{author}{Kunieda, E.}, \bibinfo{author}{Shigematsu, N.}, \bibinfo{year}{2020}.
\newblock \bibinfo{title}{Efficacy evaluation of 2d, 3d u-net semantic segmentation and atlas-based segmentation of normal lungs excluding the trachea and main bronchi}.
\newblock \bibinfo{journal}{Journal of radiation research} \bibinfo{volume}{61}, \bibinfo{pages}{257--264}.
\bibitem[{Pedregosa et~al.(2011)Pedregosa, Varoquaux, Gramfort, Michel, Thirion, Grisel, Blondel, Prettenhofer, Weiss, Dubourg et~al.}]{pedregosa2011scikit}
\bibinfo{author}{Pedregosa, F.}, \bibinfo{author}{Varoquaux, G.}, \bibinfo{author}{Gramfort, A.}, \bibinfo{author}{Michel, V.}, \bibinfo{author}{Thirion, B.}, \bibinfo{author}{Grisel, O.}, \bibinfo{author}{Blondel, M.}, \bibinfo{author}{Prettenhofer, P.}, \bibinfo{author}{Weiss, R.}, \bibinfo{author}{Dubourg, V.}, et~al., \bibinfo{year}{2011}.
\newblock \bibinfo{title}{Scikit-learn: Machine learning in python}.
\newblock \bibinfo{journal}{Journal of machine learning research} \bibinfo{volume}{12}, \bibinfo{pages}{2825--2830}.
\bibitem[{Pluim et~al.(2003)Pluim, Maintz and Viergever}]{pluim2003mutual}
\bibinfo{author}{Pluim, J.P.}, \bibinfo{author}{Maintz, J.A.}, \bibinfo{author}{Viergever, M.A.}, \bibinfo{year}{2003}.
\newblock \bibinfo{title}{Mutual-information-based registration of medical images: a survey}.
\newblock \bibinfo{journal}{IEEE transactions on medical imaging} \bibinfo{volume}{22}, \bibinfo{pages}{986--1004}.
\bibitem[{Qureshi et~al.(2001)Qureshi, Tuhrim, Broderick, Batjer, Hondo and Hanley}]{qureshi2001spontaneous}
\bibinfo{author}{Qureshi, A.I.}, \bibinfo{author}{Tuhrim, S.}, \bibinfo{author}{Broderick, J.P.}, \bibinfo{author}{Batjer, H.H.}, \bibinfo{author}{Hondo, H.}, \bibinfo{author}{Hanley, D.F.}, \bibinfo{year}{2001}.
\newblock \bibinfo{title}{Spontaneous intracerebral hemorrhage}.
\newblock \bibinfo{journal}{New England Journal of Medicine} \bibinfo{volume}{344}, \bibinfo{pages}{1450--1460}.
\bibitem[{Raghu et~al.(2019)Raghu, Zhang, Kleinberg and Bengio}]{raghu2019transfusion}
\bibinfo{author}{Raghu, M.}, \bibinfo{author}{Zhang, C.}, \bibinfo{author}{Kleinberg, J.}, \bibinfo{author}{Bengio, S.}, \bibinfo{year}{2019}.
\newblock \bibinfo{title}{Transfusion: Understanding transfer learning for medical imaging}.
\newblock \bibinfo{journal}{Advances in neural information processing systems} \bibinfo{volume}{32}.
\bibitem[{Robinson et~al.(2021)Robinson, Chuang, Sra and Jegelka}]{robinson2021contrastive}
\bibinfo{author}{Robinson, J.D.}, \bibinfo{author}{Chuang, C.Y.}, \bibinfo{author}{Sra, S.}, \bibinfo{author}{Jegelka, S.}, \bibinfo{year}{2021}.
\newblock \bibinfo{title}{Contrastive learning with hard negative samples}, in: \bibinfo{booktitle}{International Conference on Learning Representations}.
\bibitem[{Ronneberger et~al.(2015)Ronneberger, Fischer and Brox}]{ronneberger2015u}
\bibinfo{author}{Ronneberger, O.}, \bibinfo{author}{Fischer, P.}, \bibinfo{author}{Brox, T.}, \bibinfo{year}{2015}.
\newblock \bibinfo{title}{U-net: Convolutional networks for biomedical image segmentation}, in: \bibinfo{booktitle}{Medical Image Computing and Computer-Assisted Intervention--MICCAI 2015: 18th International Conference, Munich, Germany, October 5-9, 2015, Proceedings, Part III 18}, \bibinfo{organization}{Springer}. pp. \bibinfo{pages}{234--241}.
\bibitem[{Russakoff et~al.(2004)Russakoff, Tomasi, Rohlfing and Maurer}]{russakoff2004image}
\bibinfo{author}{Russakoff, D.B.}, \bibinfo{author}{Tomasi, C.}, \bibinfo{author}{Rohlfing, T.}, \bibinfo{author}{Maurer, C.R.}, \bibinfo{year}{2004}.
\newblock \bibinfo{title}{Image similarity using mutual information of regions}, in: \bibinfo{booktitle}{Computer Vision-ECCV 2004: 8th European Conference on Computer Vision, Prague, Czech Republic, May 11-14, 2004. Proceedings, Part III 8}, \bibinfo{organization}{Springer}. pp. \bibinfo{pages}{596--607}.
\bibitem[{Selvaraju et~al.(2017)Selvaraju, Cogswell, Das, Vedantam, Parikh and Batra}]{selvaraju2017grad}
\bibinfo{author}{Selvaraju, R.R.}, \bibinfo{author}{Cogswell, M.}, \bibinfo{author}{Das, A.}, \bibinfo{author}{Vedantam, R.}, \bibinfo{author}{Parikh, D.}, \bibinfo{author}{Batra, D.}, \bibinfo{year}{2017}.
\newblock \bibinfo{title}{Grad-cam: {Visual} explanations from deep networks via gradient-based localization}, in: \bibinfo{booktitle}{Proceedings of the IEEE international conference on computer vision}, pp. \bibinfo{pages}{618--626}.
\bibitem[{Studholme et~al.(1998)Studholme, Hawkes and Hill}]{studholme1998normalized}
\bibinfo{author}{Studholme, C.}, \bibinfo{author}{Hawkes, D.J.}, \bibinfo{author}{Hill, D.L.}, \bibinfo{year}{1998}.
\newblock \bibinfo{title}{Normalized entropy measure for multimodality image alignment}, in: \bibinfo{booktitle}{Medical imaging 1998: image processing}, \bibinfo{organization}{SPIE}. pp. \bibinfo{pages}{132--143}.
\bibitem[{Tang et~al.(2022)Tang, Yang, Li, Roth, Landman, Xu, Nath and Hatamizadeh}]{tang2022self}
\bibinfo{author}{Tang, Y.}, \bibinfo{author}{Yang, D.}, \bibinfo{author}{Li, W.}, \bibinfo{author}{Roth, H.R.}, \bibinfo{author}{Landman, B.}, \bibinfo{author}{Xu, D.}, \bibinfo{author}{Nath, V.}, \bibinfo{author}{Hatamizadeh, A.}, \bibinfo{year}{2022}.
\newblock \bibinfo{title}{Self-supervised pre-training of swin transformers for 3d medical image analysis}, in: \bibinfo{booktitle}{Proceedings of the IEEE/CVF Conference on Computer Vision and Pattern Recognition}, pp. \bibinfo{pages}{20730--20740}.
\bibitem[{Tian et~al.(2023)Tian, Jiang, qishuai diao, Lin, Wang and Yuan}]{tian2023designing}
\bibinfo{author}{Tian, K.}, \bibinfo{author}{Jiang, Y.}, \bibinfo{author}{qishuai diao}, \bibinfo{author}{Lin, C.}, \bibinfo{author}{Wang, L.}, \bibinfo{author}{Yuan, Z.}, \bibinfo{year}{2023}.
\newblock \bibinfo{title}{Designing {BERT} for convolutional networks: Sparse and hierarchical masked modeling}, in: \bibinfo{booktitle}{The Eleventh International Conference on Learning Representations}.
\bibitem[{Wang et~al.(2021)Wang, Shen, Yang, Lan, Xu, Wang, Zhang and Han}]{wang2021deep}
\bibinfo{author}{Wang, X.}, \bibinfo{author}{Shen, T.}, \bibinfo{author}{Yang, S.}, \bibinfo{author}{Lan, J.}, \bibinfo{author}{Xu, Y.}, \bibinfo{author}{Wang, M.}, \bibinfo{author}{Zhang, J.}, \bibinfo{author}{Han, X.}, \bibinfo{year}{2021}.
\newblock \bibinfo{title}{A deep learning algorithm for automatic detection and classification of acute intracranial hemorrhages in head ct scans}.
\newblock \bibinfo{journal}{NeuroImage: Clinical} \bibinfo{volume}{32}, \bibinfo{pages}{102785}.
\bibitem[{Wang and Bovik(2009)}]{wang2009mean}
\bibinfo{author}{Wang, Z.}, \bibinfo{author}{Bovik, A.C.}, \bibinfo{year}{2009}.
\newblock \bibinfo{title}{Mean squared error: Love it or leave it? a new look at signal fidelity measures}.
\newblock \bibinfo{journal}{IEEE signal processing magazine} \bibinfo{volume}{26}, \bibinfo{pages}{98--117}.
\bibitem[{Wang et~al.(2004)Wang, Bovik, Sheikh and Simoncelli}]{wang2004image}
\bibinfo{author}{Wang, Z.}, \bibinfo{author}{Bovik, A.C.}, \bibinfo{author}{Sheikh, H.R.}, \bibinfo{author}{Simoncelli, E.P.}, \bibinfo{year}{2004}.
\newblock \bibinfo{title}{Image quality assessment: from error visibility to structural similarity}.
\newblock \bibinfo{journal}{IEEE transactions on image processing} \bibinfo{volume}{13}, \bibinfo{pages}{600--612}.
\bibitem[{Wolf et~al.(2023)Wolf, Payer, Lisson, Lisson, Beer, G{\"o}tz and Ropinski}]{wolf2023self}
\bibinfo{author}{Wolf, D.}, \bibinfo{author}{Payer, T.}, \bibinfo{author}{Lisson, C.S.}, \bibinfo{author}{Lisson, C.G.}, \bibinfo{author}{Beer, M.}, \bibinfo{author}{G{\"o}tz, M.}, \bibinfo{author}{Ropinski, T.}, \bibinfo{year}{2023}.
\newblock \bibinfo{title}{Self-supervised pre-training with contrastive and masked autoencoder methods for dealing with small datasets in deep learning for medical imaging}.
\newblock \bibinfo{journal}{Nature Scientific Reports} \bibinfo{volume}{13}, \bibinfo{pages}{20260}.
\bibitem[{Wu et~al.(2024)Wu, Zhuang and Chen}]{wu2024voco}
\bibinfo{author}{Wu, L.}, \bibinfo{author}{Zhuang, J.}, \bibinfo{author}{Chen, H.}, \bibinfo{year}{2024}.
\newblock \bibinfo{title}{Voco: A simple-yet-effective volume contrastive learning framework for 3d medical image analysis}, in: \bibinfo{booktitle}{Proceedings of the IEEE/CVF Conference on Computer Vision and Pattern Recognition}, pp. \bibinfo{pages}{22873--22882}.
\bibitem[{Xing et~al.(2022)Xing, Huang, Nan, Wu, Wang, Gao, Walsh and Yang}]{xing2022cs}
\bibinfo{author}{Xing, X.}, \bibinfo{author}{Huang, J.}, \bibinfo{author}{Nan, Y.}, \bibinfo{author}{Wu, Y.}, \bibinfo{author}{Wang, C.}, \bibinfo{author}{Gao, Z.}, \bibinfo{author}{Walsh, S.}, \bibinfo{author}{Yang, G.}, \bibinfo{year}{2022}.
\newblock \bibinfo{title}{Cs 2: A controllable and simultaneous synthesizer of images and annotations with minimal human intervention}, in: \bibinfo{booktitle}{International Conference on Medical Image Computing and Computer-Assisted Intervention}, \bibinfo{organization}{Springer}. pp. \bibinfo{pages}{3--12}.
\bibitem[{Yang et~al.(2023)}]{medmnistv2}
\bibinfo{author}{Yang, J.}, et~al., \bibinfo{year}{2023}.
\newblock \bibinfo{title}{Medmnist v2-a large-scale lightweight benchmark for 2d and 3d biomedical image classification}.
\newblock \bibinfo{journal}{Scientific Data} \bibinfo{volume}{10}, \bibinfo{pages}{41}.
\bibitem[{Yu et~al.(2020)Yu, Yang, Wang, Leader, Wilson and Pu}]{yu20202d}
\bibinfo{author}{Yu, J.}, \bibinfo{author}{Yang, B.}, \bibinfo{author}{Wang, J.}, \bibinfo{author}{Leader, J.}, \bibinfo{author}{Wilson, D.}, \bibinfo{author}{Pu, J.}, \bibinfo{year}{2020}.
\newblock \bibinfo{title}{2d cnn versus 3d cnn for false-positive reduction in lung cancer screening}.
\newblock \bibinfo{journal}{Journal of Medical Imaging} \bibinfo{volume}{7}, \bibinfo{pages}{051202--051202}.
\bibitem[{Zettler and Mastmeyer(2021)}]{zettler2021comparison}
\bibinfo{author}{Zettler, N.}, \bibinfo{author}{Mastmeyer, A.}, \bibinfo{year}{2021}.
\newblock \bibinfo{title}{Comparison of 2d vs. 3d u-net organ segmentation in abdominal 3d ct images}, in: \bibinfo{booktitle}{International Conference on Computer Graphics, Visualization and Computer Vision 2021 - WSCG}.
\bibitem[{Zhang et~al.(2018)Zhang, Isola, Efros, Shechtman and Wang}]{zhang2018unreasonable}
\bibinfo{author}{Zhang, R.}, \bibinfo{author}{Isola, P.}, \bibinfo{author}{Efros, A.A.}, \bibinfo{author}{Shechtman, E.}, \bibinfo{author}{Wang, O.}, \bibinfo{year}{2018}.
\newblock \bibinfo{title}{The unreasonable effectiveness of deep features as a perceptual metric}, in: \bibinfo{booktitle}{Proceedings of the IEEE conference on computer vision and pattern recognition}, pp. \bibinfo{pages}{586--595}.
\bibitem[{Zhao et~al.(2020)Zhao, Zhang, He and Xie}]{yang2020covid}
\bibinfo{author}{Zhao, J.}, \bibinfo{author}{Zhang, Y.}, \bibinfo{author}{He, X.}, \bibinfo{author}{Xie, P.}, \bibinfo{year}{2020}.
\newblock \bibinfo{title}{Covid-ct-dataset: a ct scan dataset about covid-19}.
\newblock \bibinfo{journal}{arXiv preprint arXiv:2003.13865} .

\end{thebibliography}

\appendix
\section{Contrastive Learning with SwAV}\label{sec:Appendix1}
SwAV~\citep{caron2020unsupervised} starts with a large dataset of images $\underline{I} =\{I_{1}, I_{2}, I_{3}, ...\} $. For all images in one mine-batch with batch-size $Bs$, two random transformations are performed in order to obtain two randomly different images from each original image: $\underline{A} = \{A_{1}, A_{2}, A_{3}, ..., A_{Bs}  \}$ and $\underline{B} = \{B_{1}, B_{2}, B_{3}, ..., B_{Bs} \}$. The transformed images are computed by a deep learning, which can be any convolutional encoder followed by an MLP, to a latent space representation: $\underline{AQ} = \{AQ_{1}, AQ_{2}, AQ_{3}, ..., AQ_{Bs}  \}$ and $\underline{BQ} = \{BQ_{1}, BQ_{2}, BQ_{3}, ...,  BQ_{Bs} \}$. The latent space representations are further computed by feature clustering with cluster prototypes $\underline{C} = \{C_{1}, C_{2}, C_{3}, ..., C_{K}  \}$, which leads to the cluster codes
\begin{equation}
\underline{AQC} = \{\frac{1}{\tau} \cdot  \underline{AQ}^{T} \cdot \underline{C_{1}} , ...,  \frac{1}{\tau}  \cdot \underline{AQ}^{T} \cdot \underline{C_{K}} \}
\end{equation}
and 
\begin{equation}
\underline{BQC} = \{\frac{1}{\tau}  \cdot \underline{BQ}^{T} \cdot \underline{C_{1}} , ...,  \frac{1}{\tau} \cdot \underline{BQ}^{T} \cdot \underline{C_{K}} \}
\end{equation}
with the temperature value $\tau$ and the number of prototypes $K$ as hyperparameters. The model is trained to predict the cluster codes of transformed images $\underline{A}$ by the cluster codes of the transformed image $\underline{B}$ and the other way around within one min-batch by applying a cross-entropy loss with swapped predictions 
\begin{align}
L = - \sum_{k=1}^{K} \underline{BQC}_k \cdot \log{(\underline{AQC}^*_k)} 
- \sum_{k=1}^{K} \underline{BQC}_k \cdot \log{(\underline{BQC}^*_k)}, 
\end{align}
where the terms $\underline{AQC}^*_k$ and $\underline{BQC}^*_k$ are the softmax activation functions applied to the cluster codes.  

As transformations, SwAV uses color jitter, gaussian blur, and a multi-crop strategy, where two transformed images $A$ and $B$ are obtained by cropping a part of the original image with a larger crop size, and several additional samples are cropped with a smaller crop size. The transforms are implemented with torchvision with the following settings: two large crops of size 224, four small crops of the size 94, gaussian blur with probability 0.5, and color jitter with probability 0.8 and channels $(0.4,0.4,0.2,0.2)$. The cluster prototypes $\underline{C}$ are learned during training. The computed cluster codes $\underline{AQC}$ and $\underline{BQC}$ of one mini-batch should be equally partitioned by the prototypes. To ensure this equal partitioning and to avoid the trivial solution where all images collapse into the same code, the cluster codes are computed by maximizing the similarity between the latent space representations and the prototypes with the constraint
\begin{equation}
\underset{\underline{AQC}}{\textrm{max}} \: \textrm{Tr}(\underline{AQC}^T \underline{C}^T \underline{AQ}) + \epsilon H (\underline{AQ}),
\end{equation}
were $H$ is the entropy and $\epsilon$ a regularisation parameter. The same constraint for transform $B$. The clustering is performed by using the iterative Sinkhorn-Knopp algorithm~\citep{cuturi2013sinkhorn}.

Table~\ref{tab:Hyper1}, shows the hyperparameters for pre-training with SwAV. We choose the hyperparameters exactly as in the original SwAV paper. 

\begin{table}[!ht]
    \caption{Hyperparameters for pre-training with SwAV}
    \centering
    \begin{tabular}{lc}
                 \cmidrule(lr){1-2}         
        Parameters      & Values        \\
                 \cmidrule(lr){1-2} 
        Input Size      & 512      \\
        Numb. of Crops & 2; 6           \\
        Size of Crops   & 224; 96        \\
        Min Scale Crops & 0.90; 0.10    \\
        Max Scale Crops & 1.0; 0.33      \\
        Optimizer     & Lars \\
        Batch Size    & 128 \\
        Learning Rate & 0.15 \\
        Weight Decay  & 1e-6 \\
        Max Epochs    & 800 \\
        Sinkhorn Iterations  & 3\\
        Number Prototypes    & 500\\
        Freeze Prototypes    & 313\\
        Size MLP      & 2048\\
        Output Dimension    & 128\\
                 \cmidrule(lr){1-2}  
    \end{tabular}
    \label{tab:Hyper1}
\end{table}

\section{Contrastive Learning with MoCo}\label{sec:Appendix2}
MoCo~\citep{he2020momentum} starts with a large dataset of images $\{I_{1}, I_{2}, I_{3}, ...\} $, where two random transformations are performed to obtain two randomly different images from each original image: $\{A_{1}, A_{2}, A_{3}, ...  \}$ and $\{B_{1}, B_{2}, B_{3}, ...  \}$. Starting, for example, with the original image $I_{5}$, the transformed image $A_{5}$ is computed by an encoder to the latent space representation $AQ_{5}$, and the transformed image $B_{5}$ is computed by a momentum encoder to the latent space representation $BQ_{5}$. The encoders have the same architecture and can be any convolutional deep learning model. A dictionary is used to store the computed latent space representation of the momentum encoder $BQ_{5}$ together with the latent space representations of the momentum encoder from previous images $dict[..., BQ_{2}, BQ_{3}, BQ_{4}, BQ_{5}]$. The samples in the dictionary are called keys. Inside the dictionary, there is now one key that comes from the same original image as the latent space representation of the encoder. In our example, this is $BQ_{5}$ and the pair $AQ_{5}$ + $BQ_{5}$ is called positive pair. The other keys in the directory come from different original images. The pairs $AQ_{5}$ + $BQ_{4}$, $AQ_{5}$ + $BQ_{3}$, $AQ_{5}$ + $BQ_{2}$,... are called negative pairs. The model is trained to classify between positive and negative pairs by computing the InfoNCE loss
\begin{equation}
L_{10}=-\log \frac{\exp{(AQ_{5} \cdot BQ_{k} / \tau)} }{\sum_{i=0}^{5}\exp{(AQ_{5} \cdot BQ_{k} / \tau)}},
\end{equation}
which calculates a similarity score and where $\tau$ is a temperature hyperparameter. 

MoCo Version 2~\citep{chen2020improved} is an updated version of MoCo that adds an MLP projection head to the encoder and additional data transformations. As transformations, MoCo V2 uses random crop, horizontal flip, and gaussian blur. The transforms are implemented with torchvision with the following settings: two crops of size 224, gaussian blur with probability 0.5, color jitter with probability 0.8 and channels $(0.4,0.4,0.2,0.2)$, and horizontal flip with probability 0.5.

Table~\ref{tab:Hyper2}, shows the hyperparameters for pre-training with MoCo V2. We choose the hyperparameters exactly as in the original paper.

\begingroup
\setlength{\tabcolsep}{5pt}
\begin{table}[!ht]
    \caption{Hyperparamters for pre-training with MoCo}
    \centering
    \begin{tabular}{lc}
                 \toprule        
        Parameters      & Values       \\
                 \midrule
        Input Size  & 512 \\
        Number of Crops & 2 \\
        Size of Crops & 224 \\
        Optimizer     &  SGD \\
        Batch Size    & 64  \\
        Learning Rate & 1e-4 \\
         Momentum  & 0.9 \\
                 \bottomrule     
    \end{tabular}
    \label{tab:Hyper2}
\end{table}
\endgroup

\section{Masked Autoencoder with SparK}\label{sec:Appendix3}

Inspired by natural language processing, where models are pre-trained by predicting missing words in a sentence, masked autoencoders pre-train vision models by dividing the images into patches, masking some of the patches, and training the model to reconstruct the original unmasked images~\citep{ghesu2022contrastive}. SparK~\citep{tian2023designing} is the first successful adaption of masked autoencoders to Convolutional neural networks.

Starting with a large dataset of images $\{I_{1}, I_{2}, I_{3}, ...\} $, each image is divided into non-overlapping square patches and each patch is masked independently with a given probability, called ``mask ratio''. The model consists of an encoder, which can be any convolutional model and a decoder. The encoder is adapted to perform submanifold sparse convolutions, which only compute when the center of a sliding window kernel is covered by a non-masked element. The decoder is built in a U-Net~\citep{ronneberger2015u} design with three blocks of upsampling layers. The empty parts of the feature maps computed by the encoder are filled with learnable mask embeddings before being computed by the decoder. After the decoder, a head module is applied with two more upsampling layers to reach the original resolution of the input image. The model is trained with an L2 Loss between the predicted images of the model $\{I_{1}^*, I_{2}^*, I_{3}^*, ...\} $ and the original images $\{I_{1}, I_{2}, I_{3}, ...\} $, computed only on masked positions. 
For the downstream tasks, only the encoder is used. 

Table~\ref{tab:Hyper3}, shows the hyperparameters for pre-training with SparK. We choose the hyperparameters exactly as in the original paper.

\begingroup
\setlength{\tabcolsep}{4.4pt}
\begin{table}[!ht]
    \caption{Hyperparamters for pre-training with SparK}
    \centering
    \begin{tabular}{lc}
                 \toprule        
        Parameters      & Values       \\
                 \midrule
        Input Size  & $512$ \\
        Patch Site & $32\times 32$ \\
        Mask Ratio & 60\% \\
        Augmentations & horizontal flip, crop \\
        Batch Size & 32 \\
        Optimizer & LAMB \\
        Learning rate & Cosine Annealing (peak: 25e-6)\\
                \bottomrule  
    \end{tabular}
    \label{tab:Hyper3}
\end{table}
\endgroup

\section{Downstream Task Brain}\label{sec:Appendix4}
Brain hemorrhage, also known as intracranial hemorrhage, is a condition characterized by bleeding inside the skull~\cite{qureshi2001spontaneous}. Rapid diagnosis is critical because of the potential complications it can cause, including brain swelling, brain infection, or death of brain matter. The etiology of this bleeding is the rupture of blood vessels within the skull, which can be caused by factors such as physical trauma or stroke~\cite{qureshi2001spontaneous}.

An internal dataset with CT slices from 100 patients with and 100 patients without brain hemorrhage was selected by the two well-trained senior radiologists, Dr. Ch. G. Lisson and Dr. Ca. S. Lisson from the University Hospital of Ulm. Table~\ref{tab:Brain} shows details of the selected slices.

\begin{table}[!ht] 
    \caption{Downstream Task  \brain:  } 
    \centering 
    \begin{tabular}{lc}
                 \cmidrule(lr){1-2}             
        Parameters      & Values                      \\ 
                 \cmidrule(lr){1-2}   
        Format          & DICOM                            \\    
        Area            & Brain                        \\     
        Window Center   & 35/700\,HU                       \\ 
        Window Width    & 80/3020\,HU                          \\
        Tube voltage    & 100-120\,kV                 \\
        Slice Thickness & 1\,mm                       \\
        CTDI            & 33-45                         \\
        DLP             & 490-805\,mgy$\cdot$cm          \\
        Type          & No Contrast-Enhanced    \\
         Size          & 512$\times$512     \\ 
         Kernel        & Soft Tissue       \\ 
         Scanners      & PHILIPS Brilliance iCT 256            \\
                       & Siemens Somatom Definition AS+ \\
                      & Siemens Somatom Edge Plus \\
         Gender        & Unknown (anonymization) \\
         Age           & Unknown (anonymization)         \\
                 \cmidrule(lr){1-2}    
    \end{tabular}
    \label{tab:Brain}
\end{table}

\section{Accuracy of all Experiments}\label{sec:Appendix5}

\begingroup
\setlength{\tabcolsep}{40pt}
\begin{table*}[t]
    \caption{This table shows the results of the three downstream tasks \covid{}, \mnist{}, and \brain{} without using any pre-training. The weights of the model are initialized with PyTorch's standard random initialization.}
    \centering
    \resizebox{\textwidth}{!}{
    \begin{tabular}{l@{\hskip 4pt}l@{\hskip 7pt}c@{\hskip 7pt}c@{\hskip 7pt}c}
        \toprule
         \multicolumn{2}{c}{Pre-Training} & \multicolumn{3}{c}{Downstream Results}  \\
                 \cmidrule(lr){1-2} \cmidrule(lr){3-5}             
        \multicolumn{1}{c}{Dataset} & \multicolumn{1}{c}{Method} &  \covid{}  & \mnist{} & \brain{}  \\
        & &  Acc & Acc  & Acc  \\ 
        \midrule 
     - &      -                 & 0.673 $\pm$ 0.026  &   0.755 $\pm$ 0.003  & 0.596 $\pm$ 0.034 \\
        \bottomrule
    \end{tabular}
    }
    \label{tab:NoPreApp}
\end{table*}
\endgroup

\begingroup
\setlength{\tabcolsep}{40pt}
\begin{table*}[t]
    \caption{ Evaluation A: This table compares the baseline pre-training method \all{}, the current state-of-the-art, which uses all slices of a CT dataset for contrastive pre-training, with the baseline reduction pre-training method \everyn{}.
    Pre-training with SwAV is performed on the datasets \pet{} and \lidc{} with all slices, with 20\,\% of the slices by using every fifth slice, and with 10\,\% of the slices, by using every tenth slice. The different pre-trainings are evaluated on the three downstream tasks \covid{}, \mnist{}, and \brain{}. }
    \centering
    \resizebox{\textwidth}{!}{
    \begin{tabular}{l@{\hskip 4pt}l@{\hskip 7pt}c@{\hskip 7pt}c@{\hskip 7pt}c}
        \toprule
         \multicolumn{2}{c}{Pre-Training} & \multicolumn{3}{c}{Downstream Results}  \\
                 \cmidrule(lr){1-2} \cmidrule(lr){3-5}             
        \multicolumn{1}{c}{Dataset} & \multicolumn{1}{c}{Method} & \covid{}  & \mnist{} & \brain{}  \\
        & &  Acc  & Acc  & Acc  \\ 
        \midrule 
      \pet{}  & \all{}            & 0.685 $\pm$ 0.012  & 0.752 $\pm$ 0.003  & 0.628 $\pm$ 0.100 \\
              & \everyn{} 20\,\%  & 0.743 $\pm$ 0.015  & 0.789 $\pm$ 0.002  & 0.738 $\pm$ 0.047  \\
              & \everyn{} 10\,\%  & \textbf{0.755 $\pm$ 0.005}  & \textbf{0.793 $\pm$ 0.020}  & \textbf{0.772 $\pm$ 0.015} \\
      \rule{0pt}{3ex}%
      \lidc{} & \all{}            & 0.712 $\pm$ 0.015  & 0.769 $\pm$ 0.003  & 0.681 $\pm$ 0.058 \\
              & \everyn{} 20\,\%  & 0.738 $\pm$ 0.009  & 0.801 $\pm$ 0.003  & 0.681 $\pm$ 0.034 \\
              & \everyn{} 10\,\%  & \textbf{0.746 $\pm$ 0.013}  & \textbf{0.802 $\pm$ 0.002}  & \textbf{0.683 $\pm$ 0.013} \\
        \bottomrule
    \end{tabular}
    }
    \label{tab:baselineApp}
\end{table*}
\endgroup

\begingroup
\setlength{\tabcolsep}{40pt}
\begin{table*}[t]
    \caption{Evaluation B: This table compares different methods for reducing the pre-training datasets to 10\,\% of the slices. The first method is the baseline reduction method \everyn{}, which reduces the pre-training dataset by using every tenth slice, followed by the similarity based methods, which perform a pairwise comparison of all slices in a CT volume and remove one slice from pairs with high similarity.}
    \centering
    \resizebox{\textwidth}{!}{
    \begin{tabular}{l@{\hskip 4pt}l@{\hskip 7pt}c@{\hskip 7pt}c@{\hskip 7pt}c}
        \toprule
         \multicolumn{2}{c}{Pre-Training} & \multicolumn{3}{c}{Downstream Results}  \\
                 \cmidrule(lr){1-2} \cmidrule(lr){3-5}             
        \multicolumn{1}{c}{Dataset} & \multicolumn{1}{c}{Method} &  \covid{}  & \mnist{} & \brain{} \\
        & &  Acc  & Acc  & Acc  \\
        \midrule
      \pet{}   & \everyn{}   & 0.755 $\pm$ 0.005 & 0.793 $\pm$ 0.020       & 0.772 $\pm$ 0.015 \\
               & \ssim{}     & 0.752 $\pm$ 0.007  & 0.796 $\pm$ 0.002       & 0.770 $\pm$ 0.012 \\
               & \mi{}       & 0.730 $\pm$ 0.006 & 0.798 $\pm$ 0.004       & 0.772 $\pm$ 0.015 \\
               & \deep{}     & 0.712 $\pm$ 0.002 & 0.799 $\pm$ 0.003       & 0.769 $\pm$ 0.013 \\
               & \hash{}     & \textbf{0.758 $\pm$ 0.008} & \textbf{0.800 $\pm$ 0.003} & \textbf{0.775 $\pm$ 0.015} \\
          \rule{0pt}{3ex}%
      \lidc{} & \everyn{}   & 0.746 $\pm$ 0.013 & 0.802 $\pm$ 0.002       & 0.683 $\pm$ 0.013 \\
              & \ssim{}     & 0.746 $\pm$ 0.006 & 0.802 $\pm$ 0.001       & 0.758 $\pm$ 0.028 \\
              & \mi{}       & 0.748 $\pm$ 0.028 & \textbf{0.803 $\pm$ 0.004}       & 0.734 $\pm$ 0.024 \\
              & \deep{}     & 0.727 $\pm$ 0.016 & 0.801 $\pm$ 0.006       & 0.706 $\pm$ 0.054 \\
              & \hash{}     & \textbf{0.749 $\pm$ 0.009} & \textbf{0.803 $\pm$ 0.003} & \textbf{0.759 $\pm$ 0.019}  \\
        \bottomrule
    \end{tabular}
    }
    \label{tab:methodsApp}
\end{table*}
\endgroup

\begingroup
\setlength{\tabcolsep}{40pt}
\begin{table*}[t]
    \caption{Evaluation C: This table compares different similarity thresholds of the best performing reduction method \hash{}, in order to obtain the optimal degree of similarity between the slices for contrastive pre-training.}
    \centering
\resizebox{\textwidth}{!}{
    \begin{tabular}{l@{\hskip 4pt}l@{\hskip 7pt}c@{\hskip 7pt}c@{\hskip 7pt}c}
        \toprule
         \multicolumn{2}{c}{Pre-Training} & \multicolumn{3}{c}{Downstream Results}  \\
                 \cmidrule(lr){1-2} \cmidrule(lr){3-5}             
        \multicolumn{1}{c}{Dataset} & \multicolumn{1}{c}{Method} &  \covid{}  & \mnist{} & \brain{}  \\
        & &  Acc  & Acc  & Acc  \\
        \midrule 
      \pet{}  &  \hash{} - 3   & 0.749 $\pm$ 0.003  & 0.797 $\pm$ 0.002 & 0.724 $\pm$ 0.022  \\
              &  \hash{} - 6   & \textbf{0.764 $\pm$ 0.014}  & \textbf{0.799 $\pm$ 0.033} & \textbf{0.793 $\pm$ 0.022}  \\
              &  \hash{} - 12  & 0.739 $\pm$ 0.008  & 0.793 $\pm$ 0.002  & 0.703 $\pm$ 0.051  \\
             \rule{0pt}{3ex}%
      \lidc{} &  \hash{} - 3   & 0.728 $\pm$ 0.008  & 0.803 $\pm$ 0.004 & 0.752 $\pm$ 0.040  \\
              &  \hash{} - 6   & \textbf{0.755 $\pm$ 0.008}  & \textbf{0.804 $\pm$ 0.003} & \textbf{0.806 $\pm$ 0.035}  \\
              &  \hash{} - 12  & 0.726 $\pm$ 0.011  & 0.798 $\pm$ 0.003 & 0.717 $\pm$ 0.033  \\

        \bottomrule
    \end{tabular}
    }
    \label{tab:thresholdApp}
\end{table*}
\endgroup 

\begingroup
\setlength{\tabcolsep}{40pt}
\begin{table*}[t]
    \caption{Evaluation E: This table shows the downstream task results for pre-training with the contrastive learning (CL) approaches SwAV and MoCo Version 2 and the masked autoencoder (MAE) approach SparK. For all three approaches, we compare pre-training with all data (All) to pre-training with the reduced dataset using the hash reduction method and threshold 6 (Reduced) on the LIDC dataset.}
    \centering
\resizebox{\textwidth}{!}{
    \begin{tabular}{l@{\hskip 4pt}l@{\hskip 7pt}c@{\hskip 7pt}c@{\hskip 7pt}c}
        \toprule
         \multicolumn{2}{c}{Pre-Training} & \multicolumn{3}{c}{Downstream Results}  \\
                 \cmidrule(lr){1-2} \cmidrule(lr){3-5}             
        \multicolumn{1}{c}{Approach} & \multicolumn{1}{c}{Data} &  \covid{}  & \mnist{} &\brain{}  \\
        & &  Acc  & Acc  & Acc  \\
        \midrule 
      SwAV (CL)    &  All   & 0.712 $\pm$ 0.015  & 0.769 $\pm$ 0.003  & 0.681 $\pm$ 0.058 \\
              &  Reduced    & \textbf{0.755 $\pm$ 0.008}  & \textbf{0.804 $\pm$ 0.003} & \textbf{0.806 $\pm$ 0.035}  \\
             \rule{0pt}{3ex}%
      MoCoV2 (CL)   &  All   & 0.753 $\pm$ 0.014  & 0.817 $\pm$ 0.001   & 0.800 $\pm$ 0.003  \\
              &  Reduced    & \textbf{0.756 $\pm$ 0.005} & \textbf{0.819 $\pm$ 0.004} & \textbf{0.814 $\pm$ 0.032}  \\
                   \rule{0pt}{3ex}%
      SparK (MAE)  &  All   & \textbf{0.746 $\pm$ 0.005}  & \textbf{0.783 $\pm$ 0.012}  & \textbf{0.845 $\pm$ 0.003}  \\
              &  Reduced    & 0.735 $\pm$ 0.013 & 0.782 $\pm$ 0.002 & 0.841 $\pm$ 0.042  \\

        \bottomrule
    \end{tabular}
    }
    \label{tab:compareApp}
\end{table*}
\endgroup





\end{document}